\documentclass[journal]{IEEEtran}
\usepackage{amsmath}
\usepackage{graphicx}
\usepackage{multirow}
\usepackage[psamsfonts]{amssymb}
\usepackage{amsxtra}
\usepackage{threeparttable}
\usepackage{cite}
\usepackage{comment}
\usepackage{upgreek}
\usepackage{algorithm,algorithmic}
\usepackage{here}
\usepackage{color}
\usepackage{booktabs}
\usepackage{url}

\newcommand{\argmin}{\mathop{\rm argmin}\limits}
\def\x{{\mathbf x}}
\def\y{{\mathbf y}}
\def\X{{\mathbf X}}
\def\Y{{\mathbf Y}}
\def\S{{\mathbf S}}

\def\L{{\mathbf L}}

\def\D{{\mathbf D}}

\def\N{{\mathbf N}}
\def\G{{\mathcal G}}
\def\E{{\mathcal E}}
\def\V{{\mathcal V}}
\def\R{{\mathbb R}}
\def\phi{\Phi}

\def\z{{\mathbf z}}
\def\u{{\mathbf u}}

\def\v{{\mathbf v}}
\def\A{{\mathbf A}}
\def\W{{\mathbf W}}
\def\MPhi{\boldsymbol{\Upphi}}
\DeclareMathOperator{\prox}{prox}
\DeclareMathOperator{\Tr}{Tr}

\DeclareMathOperator{\psnr}{PSNR}
\DeclareMathOperator{\VEC}{vec}
\DeclareMathOperator{\RR}{\rm{R}}

\DeclareMathOperator{\DD}{\rm{D}}
\DeclareMathOperator{\PP}{\rm{P}}

\DeclareMathOperator{\sgn}{sgn}
\ifCLASSINFOpdf
\else
\fi
\begin{document}
%
\title{Robust Time-Varying Graph Signal Recovery \\for Dynamic Physical Sensor Network Data}
%
%
%

\author{Eisuke Yamagata, \textit{Student Member, IEEE}, Kazuki Naganuma, \textit{Student Member, IEEE}, Shunsuke Ono, \textit{Senior Member, IEEE}
\thanks{This work was supported by Grant-in-Aid for JSPS Fellows under Grant 23KJ0915 and 23KJ0912, in part by JST ACT-X Grant Number JPMJAX23CJ, in part by JST PRESTO under Grant JPMJPR21C4 and JST AdCORP under Grant JPMJKB2307, and in part by JSPS KAKENHI under Grant 22H03610, 22H00512, 23H01415, 23K17461, 24K03119, and 24K22291.}}
%
%

\markboth{Journal of \LaTeX\ Class Files}%
{Shell \MakeLowercase{\textit{et al.}}: Bare Demo of IEEEtran.cls for IEEE Journals}
%



\maketitle

\begin{abstract}
  We propose a time-varying graph signal recovery method for estimating the true time-varying graph signal from corrupted observations by leveraging dynamic graphs.
  Most of the conventional methods for time-varying graph signal recovery have been proposed under the assumption that the underlying graph that houses the signals is static.
  However, in light of rapid advances in sensor technology, the assumption that sensor networks are time-varying like the signals is becoming a very practical problem setting.
  In this paper, we focus on such cases and formulate dynamic graph signal recovery as a constrained convex optimization problem that simultaneously estimates both time-varying graph signals and sparsely modeled outliers. In our formulation, we use two types of regularizations, time-varying graph Laplacian-based and temporal difference-based, and also separately modeled missing values with known positions and unknown outliers to achieve robust estimations from highly degraded data. In addition, an algorithm is developed to efficiently solve the optimization problem based on a primal-dual splitting method. Extensive experiments on simulated drone remote sensing data and real-world sea surface temperature data demonstrate the advantages of the proposed method over existing methods.


\end{abstract}


%
\IEEEpeerreviewmaketitle

\section{Introduction}
%
%
%
%

\IEEEPARstart{T}{he} concept of a {\it graph signal}, defined by signal values observed on the vertex set $\V$ of a graph $\G$, has been intensely researched as an effective approach to representing irregularly structured data.
Conventional signal processing is based on spatially or temporally regular structures (e.g., images and sounds), and thus, the relations between signal values are also regular, which provides no further information for us to leverage.
On the other hand, graph signal representations and graph signal processing \cite{6409473,6494675,6557512} explicitly represent relations between signal values with vertices and weighted edges, which we can exploit as priors in the vertex domain.
Irregularly structured data such as traffic and sensor network data, geographical data, mesh data, and biomedical data all benefit from such representation.

Following the history of conventional signal processing, recovering the true graph signals from often corrupted observations is a necessity in processing the data for further use.
Many algorithms have been proposed to address the problem, most methods based on either the vertex \cite{g-tv,7179014,7979518,8901695,9415073} or the graph spectral domain \cite{7032244,7414508,9414093}.
These methods are proposed to exploit the smoothness of the signals on the vertex domain.
Graph signals are smooth when the signal values of two vertices connected by edges with large weights are similar.
Such traits can often be observed on various real-world graph signals, such as temperature data observed on a network of sensors where the edges represent the physical distances of the sensors.
Sensors that are close by would be connected by edges with large weights and would also most likely observe similar temperatures, which gives us a smooth graph signal.

Although leveraging the graph signal smoothness can successfully recover the true signals, these methods ignore the temporal domain. 
In real-life scenarios, many of the above-mentioned data can easily be sampled continuously to form time-coherent data, and therefore, priors based on the temporal domain should be effectively utilized to improve recovery results.

\subsection{Related Work}
\label{sec:related}
Several time-varying graph signal recovery methods have been discussed to leverage signal smoothness priors in the temporal domain together with that of the vertex domain. 

The earlier studies involve filtering and the Fourier transform of product graphs \cite{6879640}.
The more recent studies include spectral-based approaches like Joint time-vertex Fourier transform (JFT) \cite{7952890,7905861,8115204} 
and vertex domain-based approaches that leverage smoothness on the vertex and temporal domains \cite{7979523,9730033,9909940,10094838}.
JFT jointly applies the Fourier transform to both the temporal direction and vertex direction to fully leverage the priors of the two domains. 
In \cite{7979523,10094838}, the formulation is based on leveraging the smoothness of the temporal difference signal on the underlying graph, which effectively utilizes both domains for the signal recovery. This is generalized in \cite{9730033} by leveraging the Sobolev smoothness of the time-varying graph signals.
Low-rank-based methods \cite{liu2023time} further enhance recovery performance by leveraging the low-rankness of the time-varying graph signals.
Furthermore, distributed algorithms \cite{9146198} have been proposed to avoid costly computations like eigenvalue decomposition and matrix inversion.
Other than optimization-based methods, graph neural network-based methods \cite{10096168,10496245} have also been proposed as a more data-driven approach to time-varying graph signal recovery.

\renewcommand{\arraystretch}{1.4}
\begin{table*}[t]
  \caption{The Feature of Time-Varying Graph Signal Recovery Methods}
  \label{tb:conventional}
  \centering 
  \begin{tabular}{lcccccccc}
    \toprule 
    \multirow{2}{*}{Method} & Spatial & \multicolumn{2}{c}{Temporal correlations} & Sparse & \multirow{2}{*}{Topology} & Temporal signal & \multirow{2}{*}{Data fidelity} & \multirow{2}{*}{Algorithm} \\
    \cmidrule(lr){3-4} 
    & correlations & Graph & Signal & outliers & & regularization & & \\
    \midrule 
    JFT \cite{8115204} & $\bigcirc$ & $\times$ & $\bigcirc$ & $\times$ & Known & (Spectral Filtering) & (Spectral Filtering) & Off-line \\
    TGSR \cite{7979523} & $\bigcirc$ & $\times$ & $\bigcirc$ & $\times$ & Known & Quadratic & Cost Function & Off-line \\
    GTRSS \cite{9730033} & $\bigcirc$ & $\times$ & $\bigcirc$ & $\times$ & Known & Quadratic & Cost Function & Off-line \\
    LRGTS \cite{liu2023time} & $\bigcirc$ & $\times$ & $\bigcirc$ & $\times$ & Known & Quadratic & Constraint & Off-line \\
    DLSRA \cite{9146198} & $\bigcirc$ & $\times$ & $\bigcirc$ & $\times$ & Known & Quadratic & Cost Function & On-line \\
    DAMRA \cite{9146198} & $\bigcirc$ & $\times$ & $\bigcirc$ & $\times$ & Known & $\ell_1$ & Cost Function & On-line \\
    OLTVSG \cite{9415029} & $\bigcirc$ & $\bigtriangleup$ & $\bigtriangleup$ & $\times$ & Unknown & Quadratic & Cost Function & On-line \\
    KKF \cite{7979500} & $\bigcirc$ & $\bigcirc$ & $\bigcirc$ & $\times$ & Known & Quadratic & Cost Function & On-line \\
    Proposed & $\bigcirc$ & $\bigcirc$ & $\bigcirc$ & $\bigcirc$ & Known & Quadratic or $\ell_1$ & Constraint & Off-line \\
    \bottomrule 
  \end{tabular}
\end{table*}
\renewcommand{\arraystretch}{1.0}

Generally speaking, time-varying graph signal recovery methods are often discussed under the assumption that the underlying graph is static, especially in cases of physical sensing.
This is because, in practical applications, the graph (the weights of the edges) is pre-defined heuristically, usually by a Gaussian kernel of spatial coordinates.
Whether the graph represents a geographical network, or a cranial nerve system, or any other physical network, as long as the physical coordinates of the vertices are static, the graph is also static.
Therefore, very little work has been done in the area of time-varying graph signal recovery on a dynamic graph topology.

However, in the context of graph learning \cite{Egilmez2017,8605364,kernellearn}, where graphs are learned from the graph signals, it is much more natural to expect the graphs that house time-varying graph signals to be dynamic, just like the signals \cite{8682762}. 
Under the assumption that dynamic graphs are better representations of the underlying structures that time-varying graph signals are observed on, some recovery methods have been proposed to exploit the dynamics of the graphs \cite{7979500,9415029}.

The authors in \cite{9415029} proposed a method to estimate time-varying graph signals while also estimating dynamic graphs, which are unknown, based on an online strategy to keep track of the temporal variation of both the graph and the graph signal.
In \cite{7979500}, a space-time kernel is proposed as an extension from kernel-based methods on static graphs \cite{8682979,kernelicassp2019,9414951} to leverage both domains on a graph extended in the vertex domain to represent the temporal domain.
Although these methods allow an online estimation of the graph signals, they force the following trade-offs to their formulations in order to consider online estimation:
The inability to capture the global temporal correlations in \cite{9415029} and the limitation to quadratic costs to adopt a Kalman filtering strategy in \cite{7979500}.

Outside of signal recovery, multiplex and multilayer graph signal processing have also been considered \cite{10180031,10448337} and applied in tasks such as image segmentation and graph learning. In multiplex and multilayer graph signal processing, the relation between the signals in the temporal domain is represented by edges and weights.

To summarize, most time-varying graph signal recovery methods assume static graphs and the few that do consider dynamic graphs focus on online estimation and graph learning based on the assumption that all of the dynamic graphs are not available.

\subsection{Contributions and Paper Organization}
\label{sec:contribution}

In this paper, we focus on scenarios where the entire dynamically changing graph structure can be fully exploited. 
Conventional studies on static time-varying graph signal recovery typically assume that the graph is static and available. 
However, situations where the complete dynamic graph structure is available have not been extensively studied. 

With recent advances in sensor technology, dynamic sensor networks—such as those constructed from sensor-equipped drones and smartphones—are becoming increasingly relevant in real-world applications. 
Examples include: a) smart agriculture \cite{9092136}, where drones and IoT sensors monitor crop health, soil moisture, and weather conditions to optimize resource use and increase productivity; b) traffic and infrastructure management \cite{9893814}, where sensor-equipped vehicles and drones are used in smart cities for real-time traffic management and monitoring of road conditions; and c) environmental monitoring \cite{jonca2022drone}, where drones equipped with sensors are employed to monitor air quality, assess pollution levels, and track environmental changes.
In these and other applications that handle time-varying data from physical sensor networks, appropriate dynamic graphs can often be generated using simple algorithms, such as $k$-nearest neighbors, by leveraging the time-varying spatial locations of the sensors, similar to the static problem settings.

We formulate dynamic graph signal recovery as a constrained convex optimization problem that simultaneously estimates both time-varying graph signals and sparsely modeled outliers.
The formulation involves time-varying graph Laplacian-based and temporal difference-based regularizations that leverage the signal smoothness in both vertex and temporal domains.
Although sparse outliers are rarely addressed in time-varying graph signal recovery, we consider it to be a realistic problem since physical sensors can easily be affected by local environmental factors, which can lead to outliers.
We also restore missing values, which can represent sensor malfunctions or maintenance in real-world settings, to achieve robust estimation from highly degraded data.
Data fidelity and outlier sparsity are imposed as hard constraints rather than as a part of the cost function to facilitate parameter tuning by decoupling the parameters, as has been addressed, e.g., in \cite{CSALSA,EPIpre,ono,L0}.
A primal-dual splitting (PDS) \cite{pds,vu2013splitting} method-based algorithm is developed to efficiently solve the optimization problem.

Table \ref{tb:conventional} is a simple representation of where our method lies in relation to the conventional methods. 
Although online estimation is a crucial task in some real-world applications, the global information of the signals that are disregarded for online estimation should be effectively utilized to enhance the graph signal recovery performance.

The main contributions of this paper are as follows.
\begin{itemize}
	\item We tackle an unaddressed problem setting, that is, recovering time-varying graph signals from noisy incomplete observations, possibly with outliers, when the underlying graph topology is dynamic.
	\item We establish an optimization-based recovery method that can handle both time-varying graph Laplacian-based and temporal difference-based regularizations.
	\item We design a framework for experiments on time-varying graph signal recovery over dynamic graphs based on synthetic data, which simulates signals observed using a network of sensor-loaded drones.
	\item We conduct extensive experiments not only to validate the proposed method but also for a comprehensive study of the performance of various regularization terms over both synthetic and real-world data. 
\end{itemize}

In the following sections, we first cover the preliminaries of graph signal processing in Section \ref{sec:preliminaries} and then move on to the proposed method in Section \ref{sec:proposal}. 
In Section \ref{sec:experiments}, we illustrate the experimental results both quantitively and visually
and discuss the obtained results.

In this paper, we generalized the limited problem setting in our previous work \cite{apsipa_yamagata} and formulated a new optimization problem for signal recovery. 
Unlike in \cite{apsipa_yamagata}, we further conducted extensive experiments on synthetic data for comprehensive validation and detailed discussion. We also confirmed the validity of our method using real-world data.




\section{Preliminaries}
\label{sec:preliminaries}
\subsection{Notation}
Throughout the paper, scalars, vectors, and matrices are denoted by normal (e.g. $\lambda, \epsilon$), lowercase bold (e.g. $\x, \u$), and uppercase bold (e.g. $\L, \X$) letters, respectively. 
The element on the $i$th row, $j$th column of a matrix $\X$ is denoted by $\X(i,j)$. 
The vectorization of a matrix $\X \in \R^{n \times m}$ is denoted by $\VEC(\X) = [\x_1^\top, \x_2^\top,...,\x_m^\top]^\top$, where $\x_i$ is the $i$th column of the matrix $\X$. 
The $j$th element of a vector $\x$ is denoted by $\x(j)$, $x_j$, or $[\x]_j$. 

\subsection{Graph Laplacian}
Generally speaking, in graph signal processing, a weighted graph $\G(\V,\E,\W)$ is represented by a graph Laplacian $\L \in \R^{n \times n}$, where $\V$, $\E$, and $\W \in \R^{n \times n}$ denote the set of $n$ vertices, the set of edges, and the weight adjacency matrix, respectively. 
The combinatorial graph Laplacian of $\G$ is defined by
\begin{align}
  \L := \D - \W.
\end{align}
For the weight adjacency matrix $\W$, $\W(i,j) > 0$ when vertices $i$ and $j$ are connected, and $\W(i,j) = 0$, otherwise.
The degree matrix $\D \in \R^{n \times n}$ is a diagonal matrix whose $i$th element is the degree of the $i$th vertex. 
A degree of a vertex is the sum of the weights of all the edges connected to the vertex.
Note that all graph Laplacians of undirected graphs are real symmetric positive semidefinite matrices. 

\subsection{Smooth Graph Signal Recovery}
A graph signal $\x \in \R^{n}$ is smooth on the vertex domain when the signal values of two vertices connected by edges with large weights are similar.
The smoothness of graph signals can quantitively be measured by the graph Laplacian quadric form:
\begin{align}
  \x^\top\L\x = \sum_{i,j\in\E}\W(i,j)(\x(i) - \x(j))^2.
\end{align}
The smaller the value of the quadratic form, the smoother the graph signal $\x$ is on the graph Laplacian $\L$. 
To expand this to a series of $m$ graph signals $\X \in \R^{n \times m}$, we define the signal smoothness by
\begin{align}
  \sum_{t=1}^m \x_t^\top\L\x_t = \Tr{(\X^\top\L\X)},
\end{align}
where $\Tr{(\cdot)}$ denotes the trace of a matrix.

In the literature on optimization-based signal recovery, the graph Laplacian quadratic form is minimized to leverage signal smoothness.
Typically, the recovery of a time-varying graph signal $\overline{\X}$ from an observation $\Y = \phi(\overline{\X} + \N$) ($\N$ is some additive noise and $\phi$ is a masking operator) can be expressed as
\begin{align}
  \label{eq:trace}
  \min_{\X \in \R^{n \times m}} \Tr{(\X^\top\L\X)} + \frac{\alpha}{2}\|\X-\Y\|_F^2,
\end{align}
where $\|\cdot\|_F$ denotes the Frobenius-norm of a matrix. The second regularization term is the quadratic data fidelity term and $\alpha$ is the balancing weight. 

In addition to the graph Laplacian quadratic form, graph total variation
\begin{align}
  \mathrm{TV}_G(\x) = \sum_{i \in \E}\sqrt{\sum_{j\in\E} (\x(j)-\x(i))^2\W(i,j)^2},
\end{align}
has been proposed as the graph signal version of total variation \cite{7979518}. 

\subsection{Proximal Tools}
The proximity operator of an index $\gamma > 0$ of a proper lower-semicontinuous convex function $f$ 
is defined as
\begin{eqnarray}
  \prox_{\gamma f}: \R^n \rightarrow \R^n: \x \mapsto \argmin_{\y} f(\y) + \frac{1}{2\gamma}\|\y-\x\|^2_2,
\end{eqnarray}
in \cite{moreau1962dual}.

The indicator function of a nonempty closed convex set $C$, denoted by $\iota_C$, is defined as
\begin{eqnarray}
  \iota_C(\mathbf{x}):=
  \begin{cases}
  0, &\:\mathrm{if} \ \mathbf{x} \in C,\\
  \infty,&\:\mathrm{otherwise}.
  \end{cases}
 \end{eqnarray}
 Since the function returns $\infty$ when the input vector is outside of $C$, it acts as the hard constraint represented by $C$ in minimization. The proximity operator of $\iota_C$ is the metric projection onto $C$, defined by 
 \begin{eqnarray}
 \prox_{\iota_C}(\mathbf{x}) = P_C(\mathbf{x}) := \argmin_{\y \in C} \|\y -\x\|^2.
 \end{eqnarray}

\subsection{Primal-Dual Splitting Method}
\label{sec:pds}

A primal-dual splitting method (PDS) \cite{pds,vu2013splitting}\footnote{This algorithm is a generalization of the primal-dual hybrid gradient method \cite{chambolle2011first}} can solve optimization problems in the form of
\begin{align}
  \label{eq:pds}
  \begin{gathered}
    \min_{\mathbf{u}} f_1(\mathbf{u}) + f_2(\mathbf{u}) + f_3(\mathbf{Au}),
  \end{gathered}
\end{align}
where $f_1$ is a differentiable convex function with the $\beta$-Lipschitzian gradient $\nabla f_1$ for some $\beta > 0$, $f_2$
and $f_3$ are proximable\footnote{If the proximity operators of $f$ is computable, we call $f$ proximable.}  proper lower-semicontinuous convex functions,
and $\mathbf{A} \in \R^{n\times m}$ is a matrix.
The problem is solved by the following algorithm:
\begin{eqnarray}
  \label{eq:al_pds}
  \begin{gathered}
    \begin{split}
    \mathbf{u}^{(i+1)} &= \prox_{\gamma_1 f_2}{[\mathbf{u}^{(i)} - \gamma_1 (\nabla f_1 (\mathbf{u}^{(i)}) + \mathbf{A}^\top \mathbf{v}^{(i)})]},\\
    \mathbf{v}^{(i+1)} &= \prox_{\gamma_2 f_{3}^{*}}{[\mathbf{v}^{(i)} + \gamma_2 \mathbf{A}(2 \mathbf{u}^{(i+1)} - \mathbf{u}^{(i)})]},
    \end{split}
  \end{gathered}
\end{eqnarray}
where $f_{3}^{*}$ is the Fenchel-Rockafellar conjugate function of $f_3$ and the stepsizes $\gamma_1,\gamma_2>0$ satisfy $\frac{1}{\gamma_1} - \gamma_2 \lambda_1 (\mathbf{A}^\top\mathbf{A}) \geq \frac{\beta}{2}$ ($\lambda_1(\cdot)$ is the maximum eigenvalue of $\cdot$).
The proximity operator of $f^{*}$ can be computed as follows:
\begin{eqnarray}
  \prox_{\gamma f^{*}}(\x) = \x - \gamma \prox_{\gamma^{-1}f}(\gamma^{-1}\x).
\end{eqnarray}
The sequence generated by (\ref{eq:al_pds}) converges to a solution of (\ref{eq:pds}) under some conditions on $f_2$, $f_3$, and $\mathbf{A}$.
PDS has played a central role in various signal recovery methods, e.g., \cite{condat2014generic,ono2014hierarchical,ono2017primal,boulanger2018nonsmooth,kyochi2021epigraphical}. A comprehensive review on PDS can be found in \cite{komodakis2015playing}.

\section{Proposed Method}
\label{sec:proposal}

\subsection{Problem Formulation}
Consider the following graph signal observation model:
\begin{align}
  \label{eq:problem}
  \Y= \phi(\overline{\X} + \overline{\S} + \N),
\end{align}
where $\Y = [\y_1, \y_2,..., \y_p] \in \R^{n \times p}$ is the observation of an $n$ vertices $\times$ $p$ time-slots time-varying graph signal,
and $\overline\X \in \R^{n \times p}$ is the true signal.
For the noises, $\overline{\S} \in \R^{n \times p}$ and $\N \in \R^{n \times p}$ are sparse outliers and some random additive noise, respectively.
Missing values are represented by a masking operator $\phi$, where $\phi(\X)$ projects the elements of $\X$ to $0$ at a given probability (random vertices are masked every time-slot).
Each graph signal $\y_k$ is observed on a graph corresponding to the $k$th graph Laplacian $\L_k$.

We propose the following optimization problem to estimate the true graph signal in the above model, under the assumption that the time-varying graph signals are smooth in the vertex domain, and that the outlier $\overline{\S}$ is sparse:
\begin{align}
  \label{eq:ours}
  \begin{gathered} 
   \min_{\X\in \R^{n \times p}, \S\in \R^{n \times p}} \sum_{k=1}^p R_v^{(k)}{(\PP(\X))} + \lambda R_t(\DD(\X))\\
   \mbox{s.t.} \: \|\Y-\phi(\X+\S)\|_F \leq \varepsilon,\: \|\S\|_1 \leq \eta,
  \end{gathered}
   \end{align}
 where $R_v$ and $R_t$ indicate the graph Laplacian-based and temporal difference-based regularizations, respectively. Also, $ \X = [\x_1, \x_2, ..., \x_p] \in \R^{n \times p}$ is the estimated time-varying graph signal, $\PP(\X) := \X \:\mathrm{or}\: \DD(\X)$, $\S \in \R^{n \times p}$ is the matrix of estimated outliers, $\L_k \in \R^{n \times n}$ is the graph Laplacian at time $k$, and $\lambda > 0$ is the temporal regularization parameter.
 The $\ell_1$-norm of a matrix is denoted by $\|\cdot\|_1$, and $\varepsilon$ and $\eta$ are the radii of the Frobenius and $\ell_1$-norm balls, respectively.
 The temporal difference operator $\DD$ is defined by
 \begin{align}
  \DD(\X) = [\x_2 - \x_1, \x_3 - \x_2,..., \x_p - \x_{p-1}].
 \end{align}

 As for the first regularization term, which enforces the signal smoothness in the vertex domain, $R_v^{(k)}(\PP(\X))$ is defined as
 \begin{eqnarray}
  \label{eq:regularization}
  R_v^{(k)}{(\PP(\X))}:= \PP(\X)_k^\top\L_k\PP(\X)_k.
 \end{eqnarray}
 The above equation generalizes the following two regularizations:
\begin{align}
    R_v^{(k)}{(\X)} &= \x_k^\top\L_k\x_k,\\
   R_v^{(k)}{(\DD(\X))} &= (\x_{k+1} - \x_k)^\top\L_k(\x_{k+1} - \x_k).
\end{align}
 Graph signal smoothness is leveraged by penalizing the above two graph Laplacian quadratic forms.
  Unlike many conventional works \cite{7117446,7979523} that measure the smoothness of the time-varying graph signal on a single graph (refer Eq. \eqref{eq:trace}), the proposed regularization measures the signal smoothness per time-slot ($\sum_{k=1}^p R_v^{(k)}{(\PP(\X))}$).
 Such time-varying graph Laplacian-based regularization allows us to integrate the dynamics of the vertex domain into the formulation.
 An alternative regularization $R_v^{(k)}{(\DD(\X))}$ leverages the smoothness of the temporal difference signal on the vertex domain, proposed in \cite{7979523} under the assumption that real-world time-varying graph signals are not as smooth over the graph as expected.
On the contrary to \cite{7979523}, we consider a dynamic graph, which can question the smoothness of $\x_{k+1} - \x_k$ on $\L_k$\footnote{It should be also noted that although we are calculating $\x_{k+1} - \x_k$ on $\L_k$, there is no essential difference to calculating $\x_{k} - \x_{k-1}$ on $\L_k$ instead. If formulation applicable to online settings is desired, the latter can be used.}. However, assuming that the graph changes smoothly over time, the difference between $\L_{K+1}$ and $\L_k$ would be small, and the regularization would be effective.

 The main points of the proposed method are as follows:
 \begin{itemize}
 \item The time-varying graph Laplacian-based regularization uses a different graph Laplacian for every time-slot, unlike the conventional static-based methods.
 \item We allow the choice between two types of graph Laplacian-based regularizations.
 \end{itemize}
  
 The temporal difference-based regularization $R_t(\DD(\X))$ penalizes the temporal gradient of the graph signal $\X$ to leverage the signal smoothness in the temporal domain.
 We consider $R_t(\cdot)$ to be either $\|\cdot\|_1$ or $\|\cdot\|^2_F$ depending on the nature of the signal of interest.

 We impose data fidelity and noise sparsity as hard constraints in (\ref{eq:ours}) to facilitate parameter tuning by decoupling the parameters, as has been addressed, e.g., in \cite{CSALSA,EPIpre,ono,L0}.

 The main philosophy behind our formulation is that the underlying graph is often available in many real-world graph signal recovery situations, in which case our method would perform advantageously compared to the conventional methods \cite{7979500,9415029} that consider graph learning and online estimations.
 Especially in physical sensing situations, (e.g. using a dynamic network of sensor-loaded drones for physical sensing) the underlying dynamic graph topology can easily be generated per time-slot using the $k$-nearest neighbor algorithm and weights defined by the Gaussian kernel of spatial coordinates. 
 Therefore, the conditions where our method can recover the graph signals well are when the target distribution that is being sampled is spatially and temporally smooth. The spatial smoothness is necessary for the graph constructed by the $k$-nearest neighbor algorithm to be a decent substitute for an ideal graph. The temporal smoothness is leveraged by $R_t(\cdot)$.

 In cases where no dynamic graphs are available in any way (when $\L_1 = \L_2 = ... =\L_k$), our formulation would be the equivalent of the following formulation:
 \begin{align}
  \label{eq:ours_static}
  \begin{gathered} 
   \min_{\X\in \R^{n \times p}, \S\in \R^{n \times p}} R_v{(\PP(\X))} + \lambda R_t(\DD(\X))\\
   \mbox{s.t.}\: \|\Y-\phi(\X+\S)\|_F \leq \varepsilon,\: \|\S\|_1 \leq \eta,
  \end{gathered}
   \end{align}
   where $R_v(\PP(\X)) := \Tr{(\PP(\X)^\top\L\PP(\X))}$.
  Note that this is still a novel formulation for time-varying graph signal recovery on static graphs in terms of temporal difference-based regularization and sparse noise estimation. 

\subsection{Algorithm}
We use the primal-dual splitting method (PDS) \cite{pds} to solve (\ref{eq:ours}).
By vectorizing $\Y$ and
using indicator functions $\iota_{B_{\ell2}^{(\y,\varepsilon)}}$ and $\iota_{B_{\ell1}^{(\eta)}}$  of
\begin{align}
  B_{\ell2}^{(\y,\varepsilon)} &:= \{\z \in \R^{np}\mid \|\z - \y\|_2 \leq \varepsilon\},\\
  B_{\ell1}^{(\eta)} &:= \{\z \in \R^{np} \mid \|\z\|_1 \leq \eta\},
\end{align}
we can reformulate (\ref{eq:ours}) as the following equivalent problem:
\begin{align}
  \label{eq:pds1}
  \begin{gathered} 
  \min_{\X\in \R^{n \times p}, \S\in \R^{n \times p}} R_v{(\VEC(\PP(\X)))} + \lambda R_t{(\D\VEC{(\X)})}\\ + \iota_{B_{\ell2}^{(\VEC{(\Y)},\varepsilon)}}(\MPhi(\VEC{(\X)} + \VEC{(\S)})) + \iota_{B_{\ell1}^{(\eta)}}(\VEC{(\S)}),
  \end{gathered}
   \end{align}
   where $R_v(\x) = \x^\top\L\x = \x^\top \mathrm{diag}(\L_1, \cdots,\L_p)\x$.
  
   The matrix $\D \in \R^{n(p-1) \times np}$ and diagonal matrix $\MPhi \in \R^{np \times np}$ are the temporal linear difference operator and a masking operator for vectorized variables, respectively.

By defining $\u := [\VEC{(\X)}^\top \VEC{(\S)}^\top]^\top$, and $\v := [\v_1^\top\: \v_2^\top]^\top$ ($\v_1 \in \R^{n(p-1)}, \v_2 \in \R^{np}$), and $f_1, f_2, f_3, \A$ as
\begin{eqnarray} 
  \begin{gathered}
  \begin{split}
  \label{eq:pds2}
  f_1(\u) &:= R_v(\VEC(\PP(\X))) \\
          &=R_v\circ \VEC \circ \PP(\X),\\
  f_2(\u) &:= \iota_{B_{\ell1}^{(\eta)}}(\VEC{(\S)}),\\
  f_3(\v) &:= \lambda R_t{(\v_1)} + \iota_{B_{\ell2}^{(\VEC{(\Y)},\varepsilon)}}(\v_2),\\
  \A & := \begin{bmatrix}
          \D& \mathbf{O}\\
          \MPhi&  \MPhi \\
          \end{bmatrix},
  \end{split}
  \end{gathered}
  \end{eqnarray}
since $f_1(\u)$ is differentiable and the gradient is Lipschitz continuous and $\iota_{B_{\ell1}^{(\eta)}}(\VEC{(\S)})$ and $\lambda R_t{(\v_1)} + \iota_{B_{\ell2}^{(\VEC{(\Y)},\varepsilon)}}(\v_2)$ are proper lower-semicontinuous convex functions,
the problem in (\ref{eq:pds1}) is reduced to (\ref{eq:pds}), which can be solved by PDS (refer to Algorithm \ref{al:ours1}). Note that although the algorithm is written in the vectorized form, the actual implementation is written in the matrix form.

  The proximity operators of $f_2$ and $f_3$ can be computed by the proximity operators of $\iota_{B_{\ell1}^{(\eta)}}$, $\lambda R_t{(\v_1)}$ and $\iota_{B_{\ell2}^{(\VEC{(\Y)},\varepsilon)}}$.

The proximity operator of $f_2$ is given by
\begin{align}
  \prox_{\gamma\iota_{B_{\ell1}^{(\eta)}}}(\z) & = P_{B_{\ell1}^{(\eta)}}(\z) = \z - \prox_{\eta\|\cdot\|_\infty}(\z) \\
  &= \z - [\mathrm{sgn}(z_j)\min\{|z_j|,s\}]_{1 \leq j \leq J},
\end{align}
where $[x_j]_{1 \leq j \leq J}$ is a vector consisting $x_j$ as the $j$th element and $s \in \R$ is such that $\sum_{j=1}^J \max\{0,|z_j|-s\} = \gamma$ (Eq. (6) in \cite{ll1}).

For $f_3$, when $ R_t{}$ is $\|\cdot\|_1$, the proximity operator is equivalent to the soft-thresholding operation:
\begin{align}
  [\prox_{\gamma\|\cdot\|_1}(\z)]_j := \sgn{(z_j)}\max{\{0,|z_j|-\gamma\}},  
\end{align}
and when $ R_t{}$ is $\|\cdot\|_F^2$, the proximity operator of $f_3$ is
\begin{align}
  \prox_{\gamma\|\cdot\|_2^2}(\z) = \frac{\z}{2\gamma+1}.
\end{align}

The proximity operator of $\iota_{B_{\ell2}^{(\y,\varepsilon)}}$ is given by
\begin{align}
  \prox_{\gamma\iota_{B_{\ell2}^{(\y,\varepsilon)}}}{(\mathbf{z})} &= P_{B_{\ell2}^{(\y,\varepsilon)}}(\mathbf{z}) 
  \\ &= 
  \begin{cases}
    {}
    \mathbf{z},\;&\mathrm{if}\:\mathbf{z} \in B_{\ell2}^{(\y,\varepsilon)},\\
    \mathbf{y} + \frac{\varepsilon(\mathbf{z}-\mathbf{y})}{\|\mathbf{z}-\mathbf{y}\|_2},\;&\mathrm{otherwise}.
  \end{cases}
  \end{align}

 \subsection{Computational Complexity}
 In this section, we discuss the computational complexity per iteration of the proposed algorithm. 
 The operations in our PDS-based algorithm (Algorithm \ref{al:ours1}) that potentially require complex computations are the three proximity operations and $\nabla R_v^{(k)}$ involved in line 2 of the algorithm. The two proximity operations in lines 6 and 7 can be computed in the order of $O(np)$. For the calculation of the $\ell_1$-norm ball in line 3, it requires an order of $O(np\log np)$. 
 For $\nabla R_v^{(k)}$, the calculation of the gradient has to be performed for all time-slots, therefore, requires a computational complexity of $O(n^2p)$. 
 However, this is dependent on how dense the graph Laplacian is. If it is sufficiently sparse, like in our case where $k$-nearest neighbor ($k=4$) is applied to construct the graph, the order is reduced to $O(np)$.
 Therefore, the overall computational complexity of our algorithm is  $O(np\log np)$ per iteration.

 \begin{algorithm}[t]
  \caption{Algorithm for solving (\ref{eq:ours})}
  \label{al:ours1}
  \begin{algorithmic}[1]
  \renewcommand{\algorithmicrequire}{\textbf{Input:}}
  \renewcommand{\algorithmicensure}{\textbf{Output:}}
  \REQUIRE Input signal $\Y$, graph Laplacian $\L_k (k = 1,2,...,p)$
  \ENSURE Output signal $\X^{(i)}$ \\
  \textbf{Initialization}: $\X^{(0)} = \Y \in \mathbb{R}^{n \times p}, \S^{(0)} = \mathbf{O}$
  \WHILE{{\it A stopping criterion is not satisfied}}
   \STATE{$\VEC{(\X)}^{(i+1)} = \VEC{(\X)}^{(i)} - \gamma_1(\nabla (R_v\circ \VEC \circ \PP)(\X) + \D^\top \v_1^{(i)} + \MPhi^\top\v_2^{(i)})$;}
   \STATE{$\VEC{(\S)}^{(i+1)} = P_{B_{\ell1}^{(\eta)}}(\VEC{(\S)}^{(i)} - \gamma_1 \MPhi^\top\v_2^{(i)})$;}
   \STATE{$\mathbf{v}_1^{(i)} \leftarrow \mathbf{v}_1^{(i)} + \gamma_2 \D (2\VEC{(\X)}^{(i+1)} - \VEC{(\X)}^{(i)})$;}
   \STATE{$\mathbf{v}_2^{(i)} \leftarrow \mathbf{v}_2^{(i)} + \gamma_2 \MPhi((2\VEC{(\X)}^{(i+1)} - \VEC{(\X)}^{(i)}) +(2\VEC{(\S)}^{(i+1)} - \VEC{(\S)}^{(i)}))$;}
   \STATE{$\v_1^{(i+1)} = \v_1^{(i)} - \gamma_2 \prox_{\frac{\lambda}{\gamma_2} R_t{}}(\frac{1}{\gamma_2} \v_1^{(i)})$;}
   \STATE{$\v_2^{(i+1)} = \v_2^{(i)} - \gamma_2 P_{B_{\ell2}^{(\VEC{(\Y)},\varepsilon)}} (\frac{1}{\gamma_2} \v_2^{(i)})$;}
   \STATE{$i \leftarrow i+1$;}
  
  \ENDWHILE
  \RETURN $\X^{(i)}$ 
  \end{algorithmic} 
  \end{algorithm}

\section{Experimental Results}
\label{sec:experiments}

\subsection{Dataset}
\subsubsection{Synthetic Dataset}
\label{sec:syndata}
\begin{figure}[t]
  \begin{center}
    \begin{tabular}{c}

      \begin{minipage}{0.5\hsize}
        \begin{center}
          \includegraphics[clip, width=\hsize]{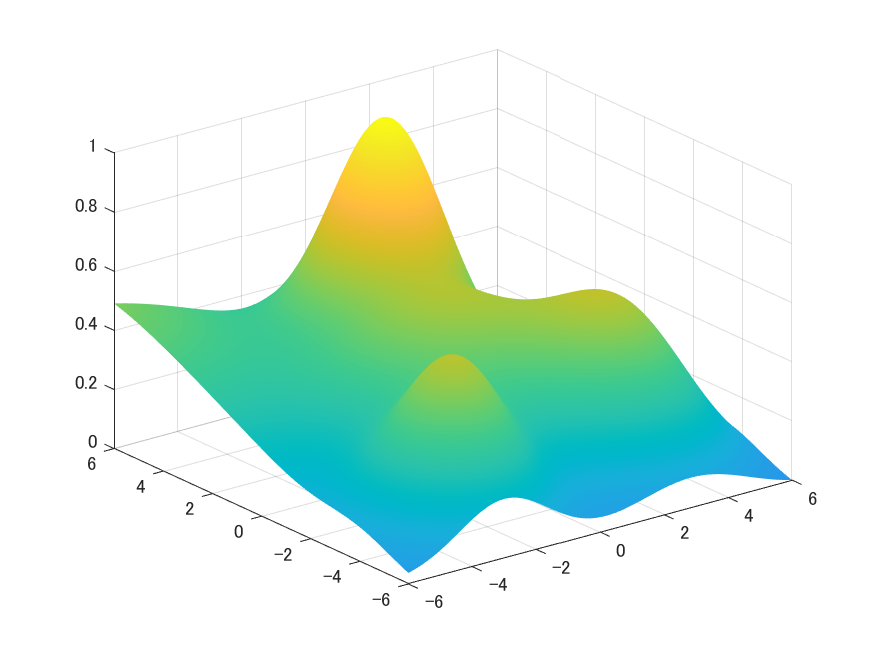}
         (a)         
        \end{center}
      \end{minipage}

      
      \begin{minipage}{0.5\hsize}
        \begin{center}
          \includegraphics[clip, width=\hsize]{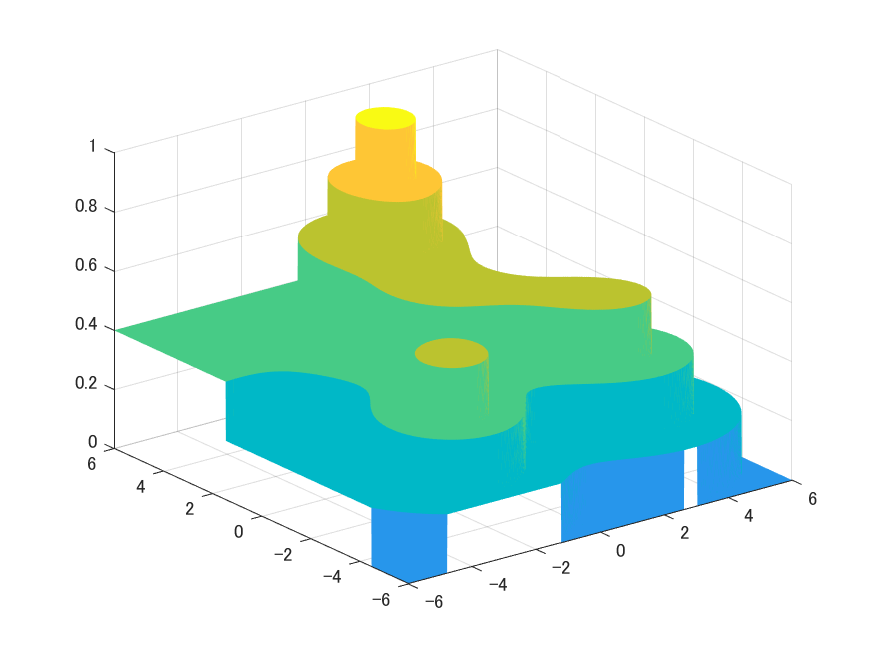} 
            
        (b)
        \end{center}
      \end{minipage}


    \end{tabular}

    \caption{(a): The smooth distribution constructed by a combination of several multivariate normal distributions. 
    (b): A piece-wise flat version of (a).}
    \vspace{-3mm}
    \label{fig:data}
  \end{center}
\end{figure}

To test the methods on a dynamic graph Laplacian setting, we constructed a synthetic dataset that simulates a network of sensor-loaded drones remotely sensing a smooth landscape. 
In a 2D plane, $128$ vertices, which represent drones, are generated randomly from a uniform distribution.
They observe signal values calculated by inputting their coordinates to the underlying smooth distribution ($[0,1]$), in this case, a combination of several multivariate normal distributions (Fig. \ref{fig:data} (a)).
The vertices move across the 2D plane at random degrees and a pre-set velocity $v$ for $100$ time-slots to generate a time-varying graph signal $\overline{\X} \in \R^{128 \times 100}$.
A dynamic graph Laplacian $\L \in \R^{128 \times 128 \times 100}$ is then constructed by applying the $k$-nearest neighbors algorithm ($k=4$) to the time-varying graph signal $\overline{\X}$.
The weights are decided by the Gaussian kernel of the spatial coordinates of the corresponding two vertices, where the variance in the Gaussian kernel is decided from the average nearest neighbor distances.

Then, $\overline{\X}$ is corrupted by adding $\overline{\S}$ and $\N_\sigma$ and being masked by $\phi$ to generate an observation $\Y$.
Outlier $\overline{\S}$ is a matrix of impulsive noise that appears at each vertex at a probability of $P_s$, where each value of the noise is uniformly distributed in the interval $[-1,1]$.
Noise $\N_\sigma$ is an additive white Gaussian noise of variance $\sigma^2$, and $\phi$ masks the signals at a probability of $P_p$ (e.g., the sampling rate is $80\%$ when $P_p = 0.2$). 
The graph construction is implemented by using GSPBox \cite{perraudin2014gspbox}.
We also constructed a piece-wise flat version of the dataset (Fig. \ref{fig:data} (b)) by rounding the signal values of Fig. \ref{fig:data} (a) to the nearest fifth of the signal range.

\subsubsection{Real-world Dataset}
\label{sec:realdata}
We use the time-varying sea surface temperature\footnote{NOAA Physical Sciences Laboratory provides the weekly global sea temperature data from their website at https://psl.noaa.gov/data/gridded/data.noaa.oisst.v2.html.} for our real-world data experiments.
The provided dataset is comprised of sea surface temperature data collected weekly at a spatial resolution of $1^\circ$ latitude $\times 1^\circ$ longitude. 
We first extract two subsets of the dataset, which originally covers a large part of the Earth's surface.
The first subset represents the South Pacific Ocean, spanning $170^\circ$ west to $90^\circ$ west and $60^\circ$ south to $10^\circ$ north.
The second subset represents the North Atlantic Ocean, spanning $60^\circ$ west to $20^\circ$ west and $50^\circ$ north to $70^\circ$ south.
Then, we randomly select $100$ and $64$ initiate points from the respective cut-out data and simulate a dynamic graph by sequentially shifting these points to adjacent positions at each timeslot.
Using this method, we create two datasets ($100$ nodes $\times$ $600$ time-slots and $64$ nodes $\times$ $300$ time-slots, respectively) similar to the synthetic dataset where the underlying graph is dynamic.
The graphs at each time-slots are constructed and the graph signals are corrupted in the same manner as the synthetic dataset.
Note that unlike the synthetic dataset, where the underlying distribution is static, the sea temperature is also time-varying.
The temperature data is scaled to $[0\: 1]$.

\begin{table*}[!htb]
  \caption{The Average Recovery Performance on The Various Datasets Measured in RMSE And MAE. The Noise Level Is Set to $\sigma = 0.1, P_s = 0.1, P_p = 0.1$. Sensor Velocity Is Set as $v=0.25$ for The Synthetic Datasets. Methods That Leverage Static Graphs Are in \textcolor{blue}{Blue} And Methods That Leverage Dynamic Graphs Are in \textcolor{red}{Red}.}
  \label{tb:result}
  \centering
  \scalebox{0.93}{
    \begin{tabular}{ccccccccccc}
      \toprule
      \multirow{2}{*}{Method} & \multirow{2}{*}{Vertex domain} & \multirow{2}{*}{Temporal domain} &\multicolumn{2}{c}{Syn. (smooth)} & \multicolumn{2}{c}{Syn. (piece-wise flat)} &\multicolumn{2}{c}{SST Pacific}&\multicolumn{2}{c}{SST Atlantic} \\
      \cmidrule(lr){4-5} \cmidrule(lr){6-7} \cmidrule(lr){8-9} \cmidrule(lr){10-11} 
      & & & RMSE & MAE & RMSE & MAE & RMSE & MAE & RMSE & MAE  \\
      \midrule
       $\mathrm{L}_1$                         & - & $\|\DD(\X)\|_1$                                      & $0.0582 $ & $0.0446 $ & $0.0614 $ & $0.0436 $ & $0.0643 $ & $0.0487 $                                             & $ 0.0796$ & $0.0635$\\
       $\mathrm{L}_2$                         & - & $\|\DD(\X)\|_F^2$                                    & $0.0507 $ & $0.0373 $ & $0.0587 $ & $0.0445 $ & $0.0516 $ & $0.0394 $                                             & $ 0.0668$ & $0.0529$\\
      \textcolor{blue}{S-TV}  \cite{7979518}  & $\mathrm{TV}_G(\X)$ & -                                  & $0.2791 $ & $0.1949 $ & $0.2807 $ & $0.1947 $ & $0.3691 $ & $0.2390 $                                             & $ 0.3486$ & $0.2307$\\
      \textcolor{blue}{S-X}  \cite{7117446}   & $R_v{(\X)}$ & -                                          & $0.1443 $ & $0.1036 $ & $0.1539 $ & $0.1104 $ & $0.1763 $ & $0.1249 $                                             & $ 0.1128$ & $0.0846$\\
      \textcolor{red}{D-X}                    & $\sum_{k=1}^{p} R_v^{(k)}{(\X)}$ & -                     & $0.0956 $ & $0.0704 $ & $0.1035 $ & $0.0772 $ & $0.0826 $ & $0.0628 $                                             & $ 0.0729$ & $0.0564$\\
      \textcolor{blue}{TGSR}  \cite{7979523}  & $R_v{(\DD(\X))}$ & -                                     & $0.0727 $ & $0.0532 $ & $0.0780 $ & $0.0588 $ & $0.0761 $ & $0.0576 $                                             & $ 0.0802$ & $0.0618$\\
      \textcolor{blue}{GTRSS}  \cite{9730033} & $\Tr(\DD(\X)^\top(\L+\epsilon\mathbf{I})^\beta\DD(\X))$ & -              & $0.0586 $ & $0.0424 $ & $0.0652 $ & $0.0492 $ & $0.0571 $ & $0.0440 $                             & $ 0.0667$ & $0.0521$\\
      \textcolor{blue}{S-DX-$\mathrm{L}_1$}   & $R_v{(\DD(\X))}$&$\|\DD(\X)\|_1$                         & $0.0530 $ & $0.0390 $ & $0.0590 $ & $0.0431 $ & $0.0538 $ & $0.0408 $                                             & $ 0.0680$ & $0.0531$\\
      \textcolor{red}{D-DX-$\mathrm{L}_1$}    & $\sum_{k=1}^{p-1}R_v^{(k)}{(\DD(\X))}$&$\|\DD(\X)\|_1$   & $0.0549 $ & $0.0404 $ & $0.0601 $ & $0.0436 $ & $0.0527 $ & $0.0405 $                                             & $ 0.0681$ & $0.0533$\\
      \textcolor{blue}{S-DX-$\mathrm{L}_2$}   & $R_v{(\DD(\X))}$&$\|\DD(\X)\|_F^2$                       & $0.0494 $ & $0.0372 $ & $0.0579 $ & $0.0440 $ & $0.0511 $ & $0.0389 $                                             & $ 0.0643$ & $0.0504$\\
      \textcolor{red}{D-DX-$\mathrm{L}_2$}    & $\sum_{k=1}^{p-1}R_v^{(k)}{(\DD(\X))}$&$\|\DD(\X)\|_F^2$ & $0.0497 $ & $0.0374 $ & $0.0579 $ & $0.0440 $ & $0.0502 $ & $0.0388 $                                             & $ 0.0650$ & $0.0511$\\
      \textcolor{blue}{S-X-$\mathrm{L}_1$}    & $R_v{(\X)}$ & $\|\DD(\X)\|_1$                            & $0.0544 $ & $0.0402 $ & $0.0620 $ & $0.0456 $ & $0.0586 $ & $0.0445 $                                             & $ 0.0718$ & $0.0573$\\
      \textcolor{red}{D-X-$\mathrm{L}_1$}     & $\sum_{k=1}^{p} R_v^{(k)}{(\X)}$ & $\|\DD(\X)\|_1$       & $0.0492 $ & $0.0365 $ & $0.0566 $ & $\mathbf{0.0410} $ & $0.0513 $ & $0.0394 $                                    & $ 0.0556$ & $0.0444$\\
      \textcolor{blue}{S-X-$\mathrm{L}_2$}    & $R_v{(\X)}$ & $\|\DD(\X)\|_F^2$                          & $0.0545 $ & $0.0403 $ & $0.0631 $ & $0.0480 $ & $0.0586 $ & $0.0445 $                                             & $ 0.0718$ & $0.0573$\\
      \textcolor{red}{D-X-$\mathrm{L}_2$}     & $\sum_{k=1}^{p} R_v^{(k)}{(\X)}$ & $\|\DD(\X)\|_F^2$     & $\mathbf{0.0453} $ & $\mathbf{0.0335} $ & $\mathbf{0.0550} $ & $0.0417 $ & $\mathbf{0.0449} $ & $\mathbf{0.0347} $& $ \mathbf{0.0519}$ & $\mathbf{0.0413}$\\
      \midrule                                                                                              
      Observ.                                &&                                                             & $0.2371 $ & $0.1482 $ & $0.2368 $ & $0.1463 $ & $0.2383 $ & $0.1482 $                                          & $ 0.2787$ & $0.0341$\\
      \bottomrule
    \end{tabular}
  }
\end{table*}

\begin{figure*}[h]
  \begin{center}
    \begin{tabular}{c}

      \begin{minipage}{0.25\hsize}
        \begin{center}
          \includegraphics[clip, width=\hsize]{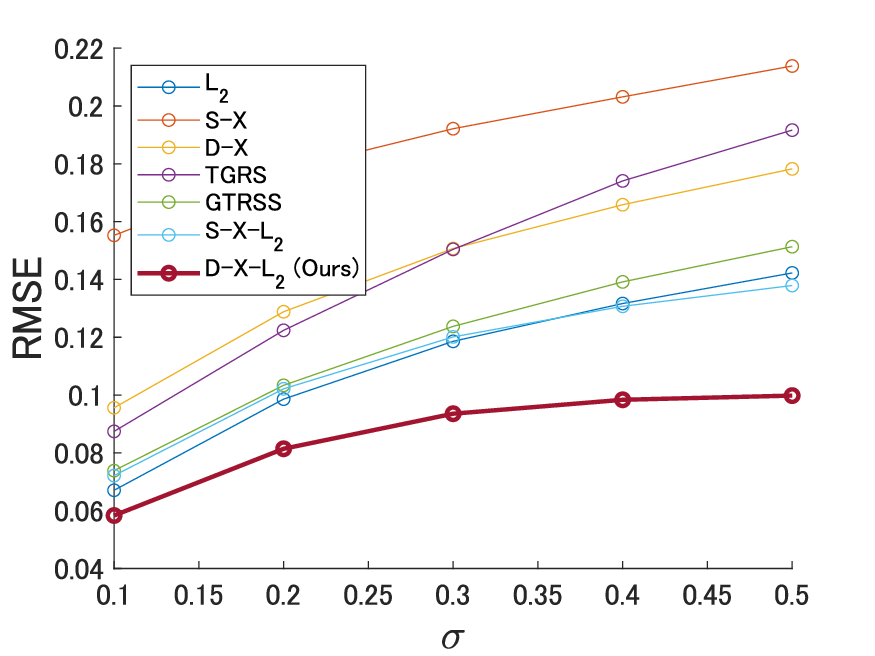}
         (a) Noise level: $\sigma$
        \end{center}
      \end{minipage}

      
      \begin{minipage}{0.25\hsize}
        \begin{center}
          \includegraphics[clip, width=\hsize]{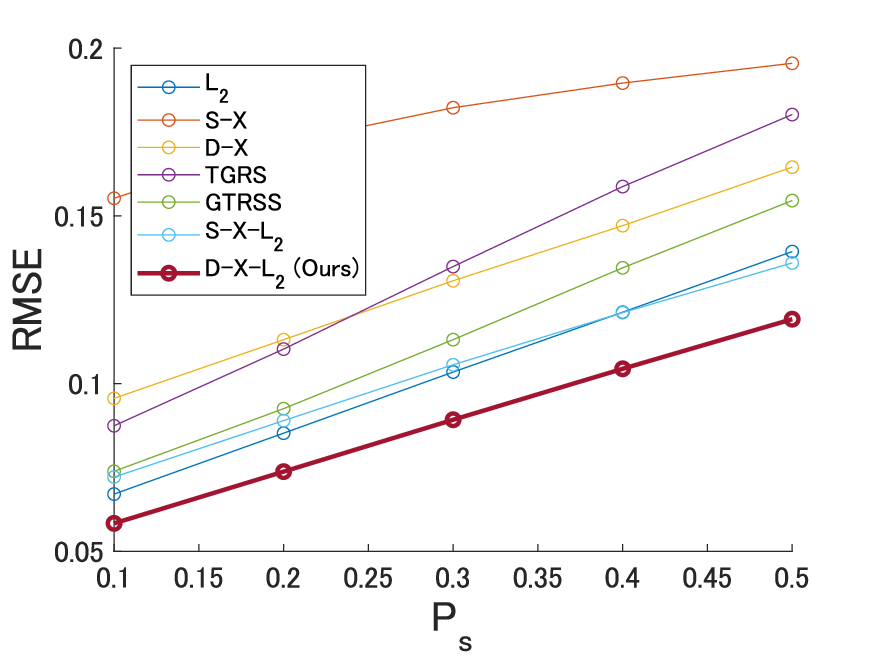} 
        (b) Noise level: $P_s$
        \end{center}
      \end{minipage}

      \begin{minipage}{0.25\hsize}
        \begin{center}
          \includegraphics[clip, width=\hsize]{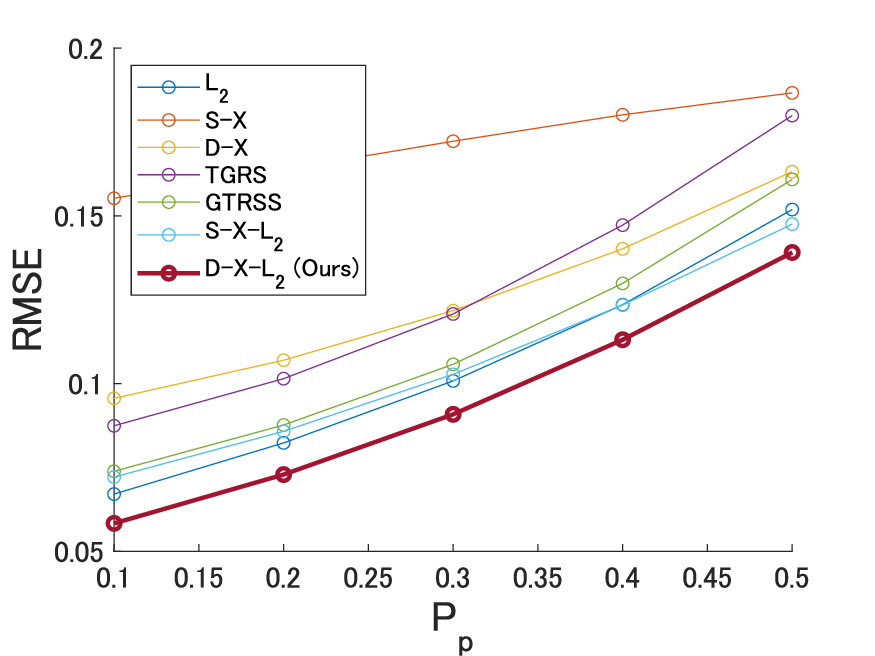} 
        (c) Noise level: $P_p$
        \end{center}
      \end{minipage}

      \begin{minipage}{0.25\hsize}
        \begin{center}
          \includegraphics[clip, width=\hsize]{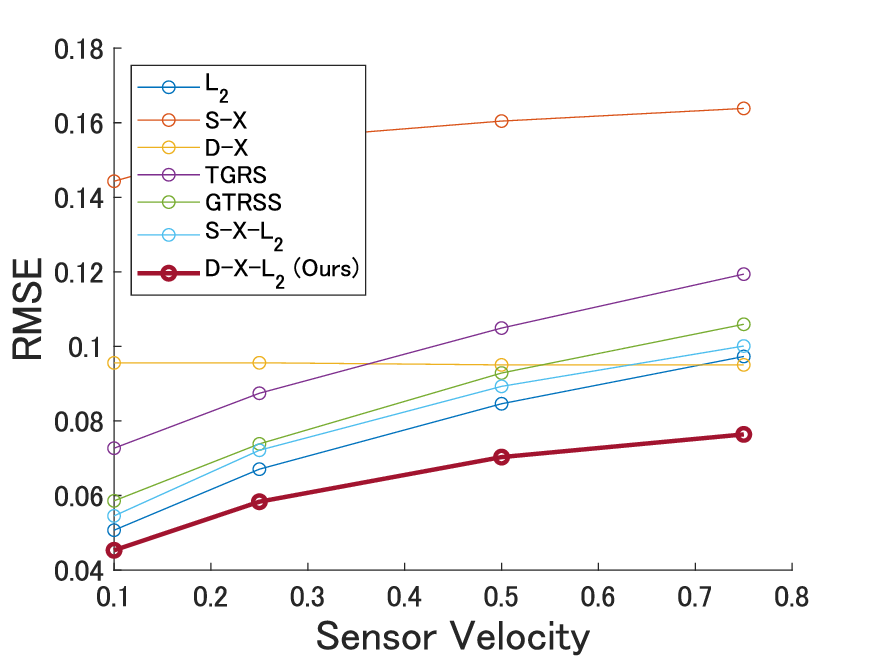} 
        (d) Sensor velocity
        \end{center}
      \end{minipage}


    \end{tabular}

    \caption{The graphs of RMSE vs various noise levels and sensor velocity on the $128\times100$ synthetic (smooth) dataset. The fixed noise levels are set as $\sigma=0.1, P_s=0.1, P_p=0.1$ and velocity $v = 0.25$.}
    \vspace{-5mm}
    \label{fig:ablation}
  \end{center}
\end{figure*}

\begin{figure}[h]
  \begin{center}

          \includegraphics[clip, width=\hsize]{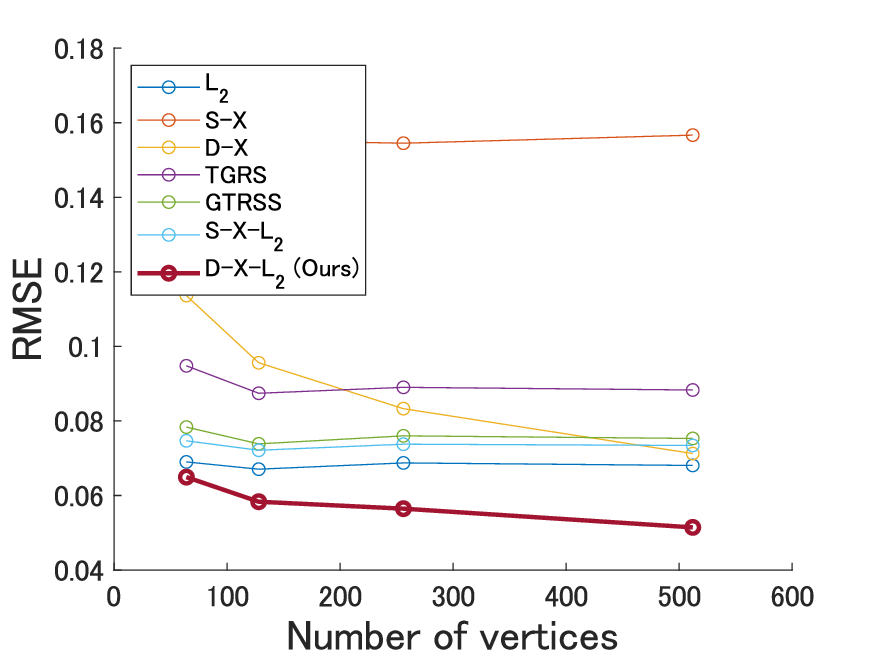}

    \caption{The graph of RMSE vs various network sizes on the synthetic (smooth) dataset. The noise levels are set as $\sigma=0.1, P_s=0.1, P_p=0.1$ and the sensor velocity is $v=0.25$.}
    \vspace{-5mm}
    \label{fig:graph_size}
  \end{center}
\end{figure}

\subsection{Experimental Settings}
\label{sec:ex_setting}
In the experiments\footnote{The codes are available at \url{https://drive.google.com/drive/folders/1RZiShedqBn2qV7y80pAoRqb3XRM-BBlF?usp=sharing}.}, we evaluate our method on synthetic and real-world datasets.
Regarding the implementation for Algorithm \ref{al:ours1}, the stopping criterion is set to $\frac{\|\X^{(i)} - \X^{(i-1)}\|_2}{\|\X^{(i-1)}\|_2} \leq 1.0 \times 10^{-4} \land \frac{\|\S^{(i)} - \S^{(i-1)}\|_2}{\|\S^{(i-1)}\|_2} \leq 1.0 \times 10^{-4}$.
The performance of the methods is measured in $\mathrm{RMSE}=\sqrt{\frac{1}{n \times p} \sum_{i=1}^{n} \sum_{j=1}^{p} (\overline{\X}(i,j) - \X(i,j))^2}$ and $\mathrm{MAE} = \frac{1}{n \times p} \sum_{i=1}^{n} \sum_{j=1}^{p} |\overline{\X}(i,j) - \X(i,j)|$.

We compare several combinations of vertex- and temporal-domain regularizations. 
The methods are labeled based on 1. whether it leverages a static or a dynamic graph (S/D-), 2. the content of $\PP(\X)$ used for the graph Laplacian-based regularization (-X/DX-), and 3. whether $\|\cdot\|_1$ or $\|\cdot\|_F^2$ is used for the temporal difference-based regularization (-$\mathrm{L}_1$/$\mathrm{L}_2$). (refer to the method column of Table \ref{tb:result}).
Note that the Methods D-DX-$\mathrm{L}_1$, D-DX-$\mathrm{L}_2$, D-X-$\mathrm{L}_1$, and D-X-$\mathrm{L}_2$ are the different implementations of the proposed formulation (\ref{eq:ours}) where S-DX-$\mathrm{L}_1$, S-DX-$\mathrm{L}_2$, S-X-$\mathrm{L}_1$, and S-X-$\mathrm{L}_2$ are their static counterparts, respectively.

Five trials are conducted for every setting and the averaged results are presented.
For all the methods that use the regularization parameter $\lambda$, we conduct a linear search for every trial. Then we determine a value that performs well for all trials on average. The results of the trials that used the determined $\lambda$ are used to give the final averaged results.
For static methods that use only one graph Laplacian, we input $\L_{p/2}$, the graph constructed at the middle of the time-slots ($t = p/2$).

We compare our methods with TGSR \cite{7979523} and GTRSS \cite{9730033}, which are state-of-the-art time-varying graph signal recovery methods that leverage signal smoothness in the vertex and temporal domains.
For a fair comparison, all methods, including the comparison methods, have been modified to support the same $\ell_1$-norm ball constraint as the proposed formulation. 

All of the methods share the same constraints $\|\Y-\phi(\X+\S)\|_F \leq \varepsilon, \|\S\|_1 \leq \eta$.
This is to avoid parameter tuning in cases where the noise level is known a priori or can be estimated, and in our experiments, $\varepsilon$ and $\eta$ are set to:
\begin{eqnarray}
  \label{eq:noiselevel}
  \varepsilon = 0.9\sigma\sqrt{np(1-P_s)(1-P_p)},\:
  \eta = \frac{P_s}{2}np.
  \end{eqnarray}
All of the methods are implemented by PDS.
The stepsizes are set as follows in correspondence to the condition mentioned in Section \ref{sec:pds}: $\gamma_1 = \frac{1}{\beta}$, $\gamma_2 = \frac{0.49}{\gamma_1\lambda_1 (\mathbf{A}^\top\mathbf{A})}$.

  \begin{figure*}[h]
    \begin{center}
      \begin{tabular}{c}
  
        \begin{minipage}{0.25\hsize}
          \begin{center}
            \includegraphics[clip, width=\hsize]{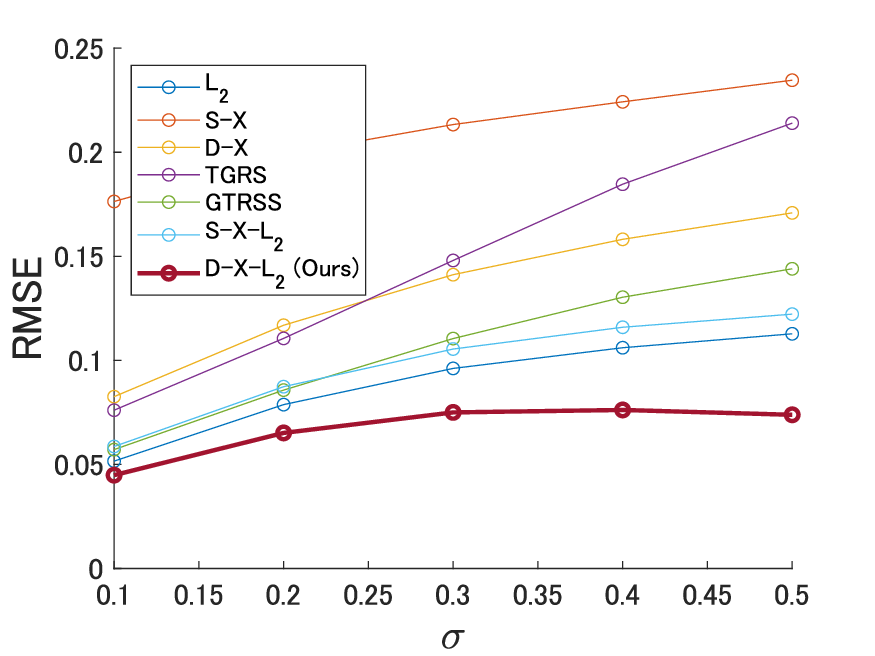}
           (a) Noise level: $\sigma$
          \end{center}
        \end{minipage}
  
        
        \begin{minipage}{0.25\hsize}
          \begin{center}
            \includegraphics[clip, width=\hsize]{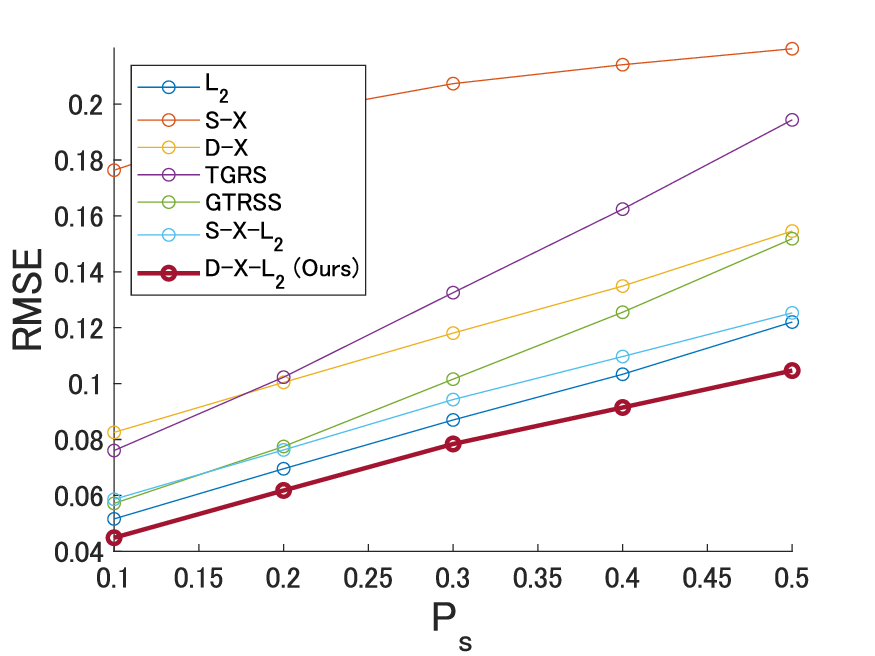} 
          (b) Noise level: $P_s$
          \end{center}
        \end{minipage}
  
        \begin{minipage}{0.25\hsize}
          \begin{center}
            \includegraphics[clip, width=\hsize]{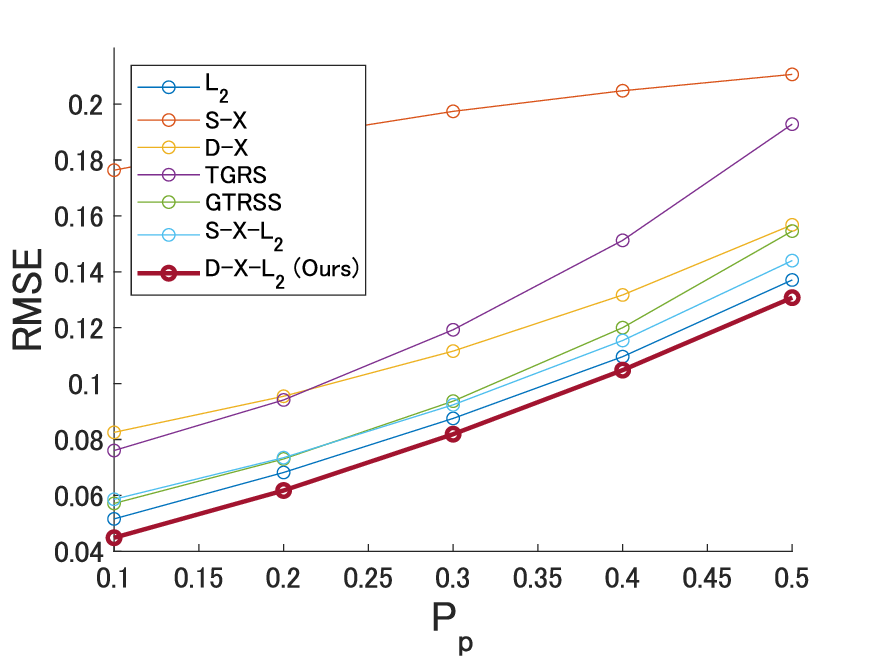} 
          (c) Noise level: $P_p$
          \end{center}
        \end{minipage}

        \begin{minipage}{0.25\hsize}
          \begin{center}
            \includegraphics[clip, width=\hsize]{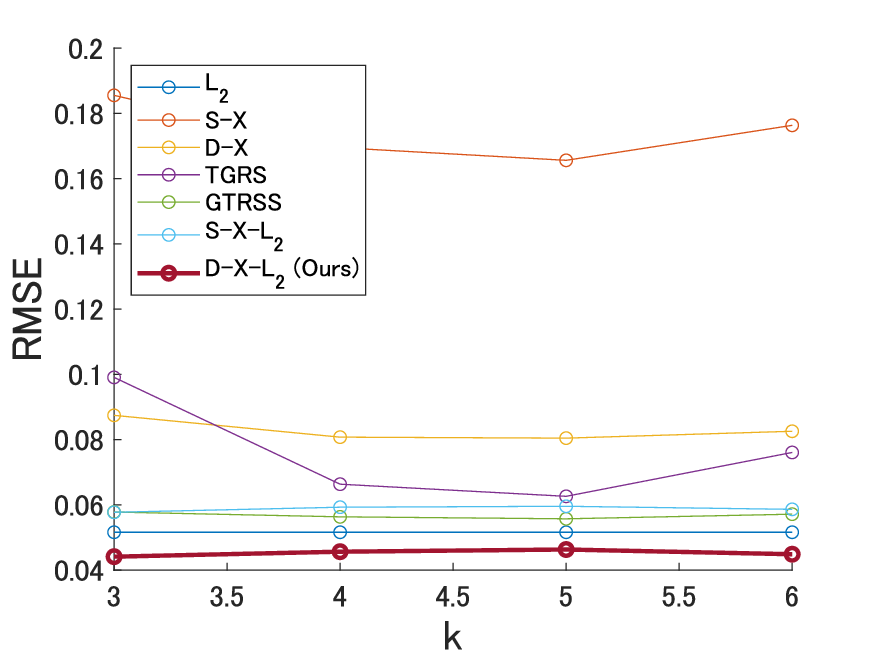} 
          (d) Num. of edges: $k$
          \end{center}
        \end{minipage}

      \end{tabular}
  
      \caption{The graphs of RMSE vs various noise levels and $k$ for graph construction on the sea surface temperature (Pacific) dataset. The fixed noise levels are set as $\sigma=0.1, P_s=0.1, P_p=0.1$, $k = 4$.}
      \vspace{-5mm}
      \label{fig:ablation_sst}
    \end{center}
  \end{figure*}

  \begin{figure}[t]
    \begin{center}
      
          \begin{center}
            \includegraphics[clip, width=\hsize]{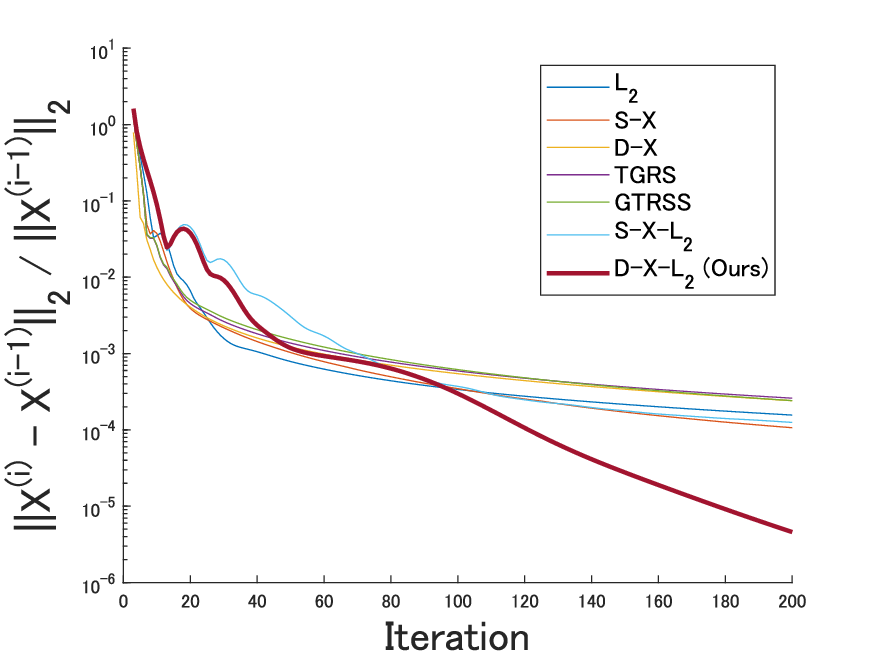} 
          \end{center}
	    
      \caption{The graph of convergence rate on the sea surface temperature dataset (Pacific). The fixed noise levels are set as $\sigma=0.1, P_s=0.1, P_p=0.1$.}
      \vspace{-5mm}
      \label{fig:convergence}
    \end{center}
  \end{figure}

  \begin{figure*}[ht]
    \begin{center}
      \begin{tabular}{c}
     
        \begin{minipage}{0.20\hsize}
          \begin{center}
            \includegraphics[width=\hsize]{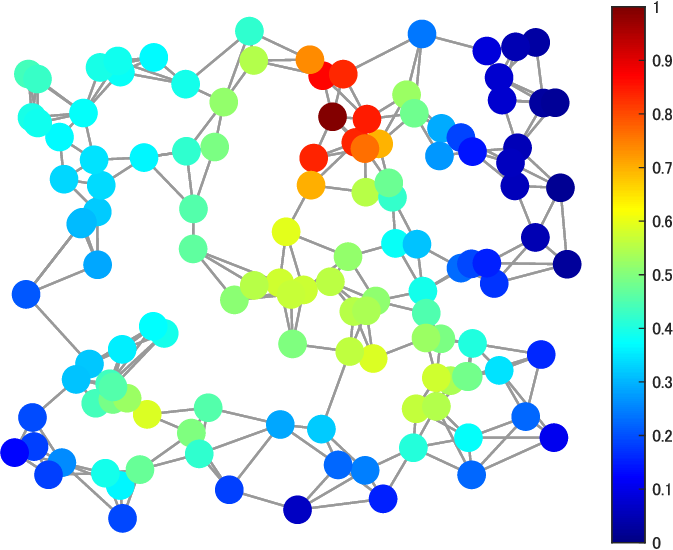}
            {\\ Original \\  $n = 128$}
            \vspace{2mm}
          \end{center}
        \end{minipage}

         \begin{minipage}{0.20\hsize}
          \begin{center}
            \includegraphics[width=\hsize]{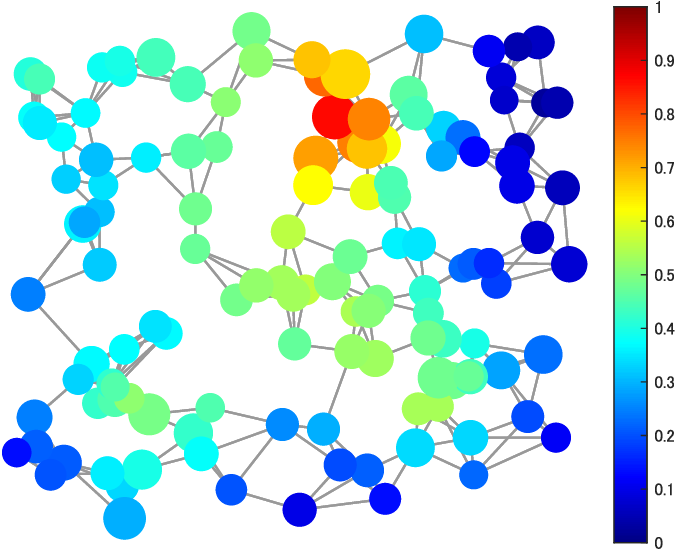}
             RMSE: $ 0.0443$ \\  $\mathrm{L}_2$
             \vspace{2mm}
          \end{center}
        \end{minipage}

         \begin{minipage}{0.20\hsize}
          \begin{center}
            \includegraphics[width=\hsize]{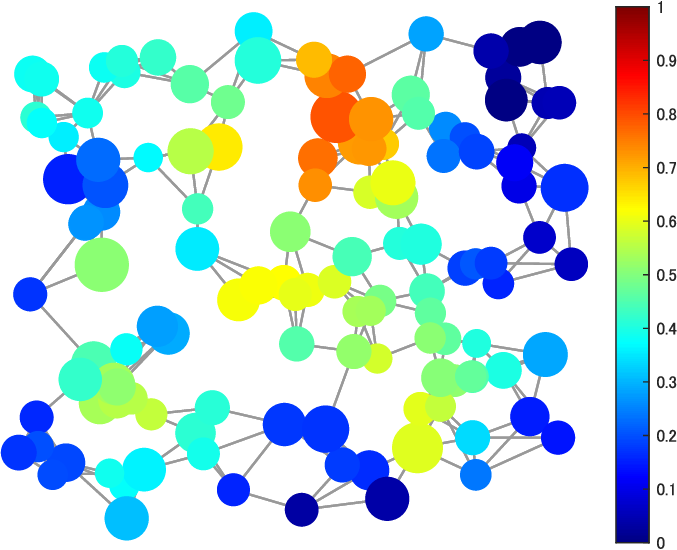}
            \\ RMSE: $ 0.1550$ \\  \textcolor{blue}{S-X} \cite{7117446}
            \vspace{2mm}
          \end{center}
        \end{minipage}
        
    
         \begin{minipage}{0.20\hsize}
          \begin{center}
            \includegraphics[width=\hsize]{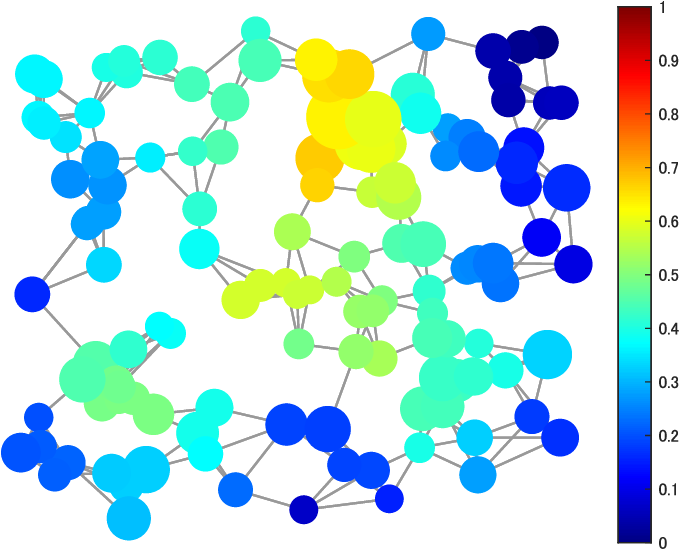}
         \\ RMSE: $ 0.0951$ \\  \textcolor{red}{D-X}
         \vspace{2mm}
          \end{center}
        \end{minipage}

        \begin{minipage}{0.20\hsize}
          \begin{center}
            \includegraphics[width=\hsize]{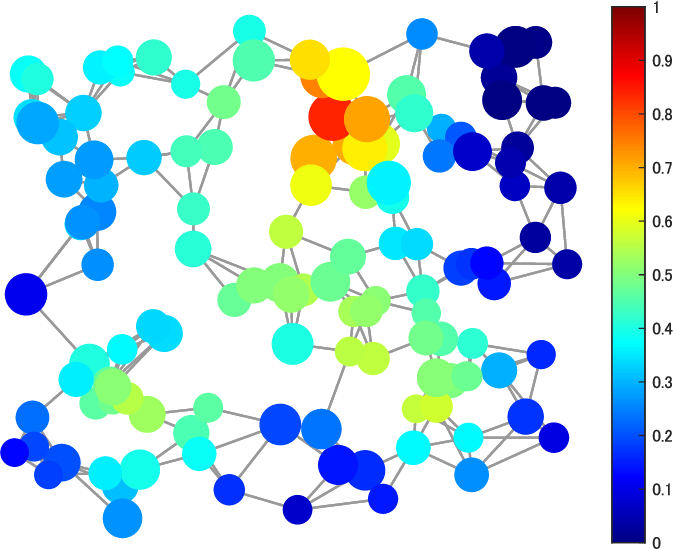}
        \\ RMSE: $ 0.0711$ \\  \textcolor{blue}{TGSR} \cite{7979523}
        \vspace{2mm}
          \end{center}
        \end{minipage}
          \\

        \hspace{-0.21\hsize}
        \begin{minipage}{0.20\hsize}
          \begin{center}
            \includegraphics[width=\hsize]{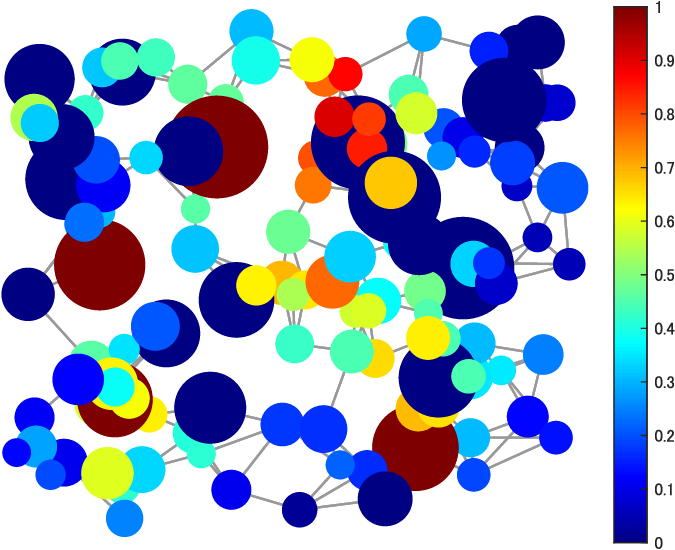}
             \\ RMSE: $ 0.2371$ \\ Obsevation
             \vspace{2mm}
          \end{center}
        \end{minipage}

        \begin{minipage}{0.20\hsize}
          \begin{center}
            \includegraphics[width=\hsize]{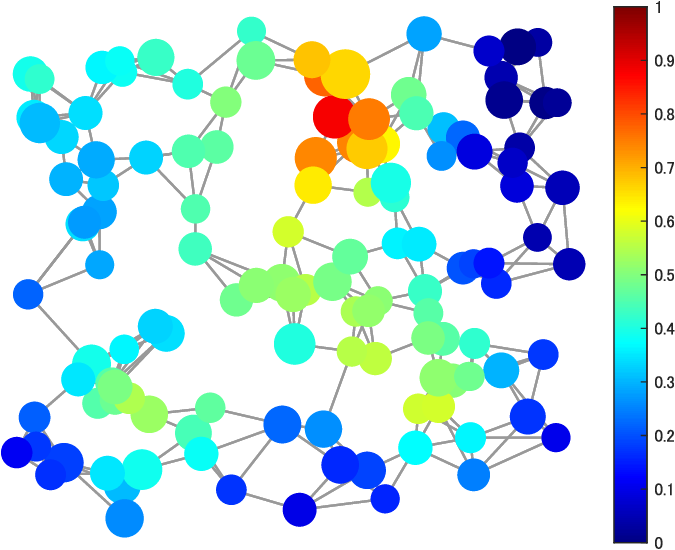}
          \\ RMSE: $ 0.0551$ \\  \textcolor{blue}{GTRSS} \cite{9730033}
          \vspace{2mm}
          \end{center}
        \end{minipage}

        \begin{minipage}{0.20\hsize}
          \begin{center}
            \includegraphics[width=\hsize]{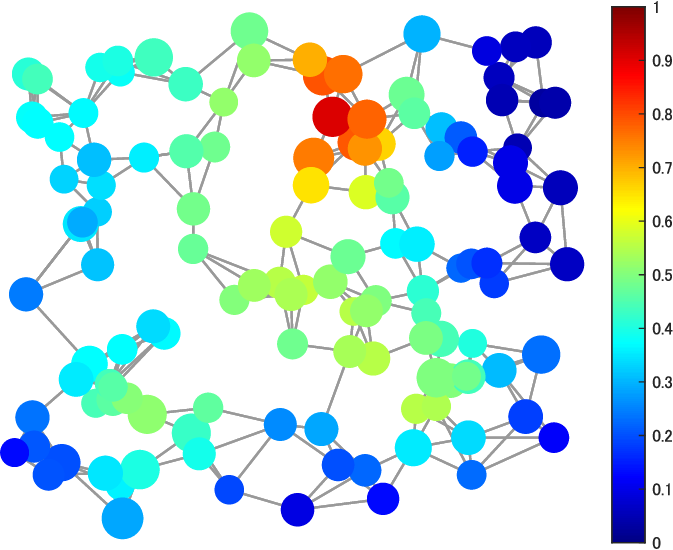}
          \\  RMSE: $ 0.0497$ \\  \textcolor{blue}{S-X-$\mathrm{L}_2$} 
          \vspace{2mm}
          \end{center}
        \end{minipage}
        
        \begin{minipage}{0.20\hsize}
          \begin{center}
            \includegraphics[width=\hsize]{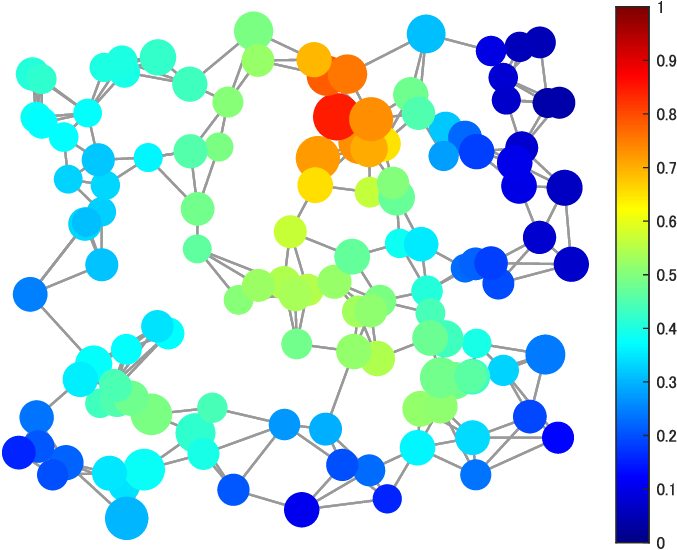}
          \\ {\bf RMSE: ${\bf 0.0383}$ \\  \textcolor{red}{D-X-$\mathrm{L}_2$} Ours}
          \vspace{2mm}
          \end{center}
        \end{minipage}

      \end{tabular}
      \caption{Graph signal recovery results of the synthetic $128\times100$ dataset ($\sigma,P_s,P_p = 0.1, v = 0.25$). The area of the nodes is proportional to the magnitude of the error, where larger areas indicate larger errors.
      Note that although the color range for the observations is clipped to [0,1] to unify the colorbars of all the figures, the magnitude of error represented by the size of the nodes is not clipped, therefore, some nodes may be colored the same but differently sized.}
      \vspace{-6mm}
      \label{fig:denoise_128_}
  
    \end{center}
  \end{figure*}

  \begin{figure*}[!htb]
    \begin{center}
      \begin{tabular}{c}
     
        \begin{minipage}{0.20\hsize}
          \begin{center}
            \includegraphics[width=\hsize]{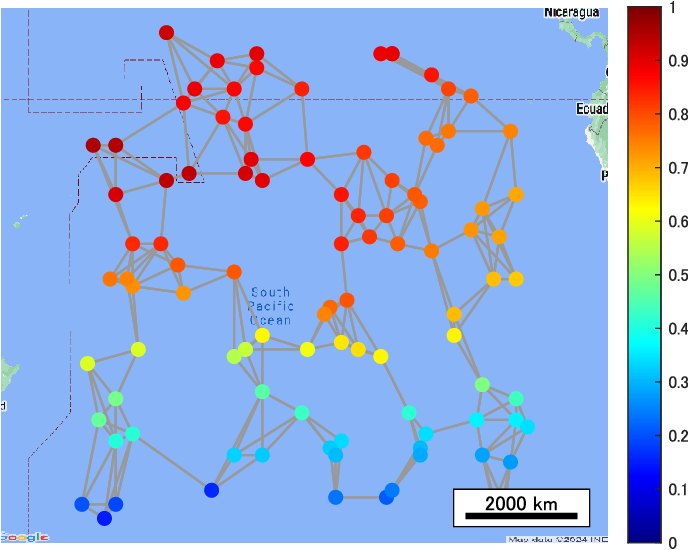}
            {\\ Original \\  $n = 100$}
            \vspace{2mm}
          \end{center}
        \end{minipage}
        


         \begin{minipage}{0.20\hsize}
          \begin{center}
            \includegraphics[width=\hsize]{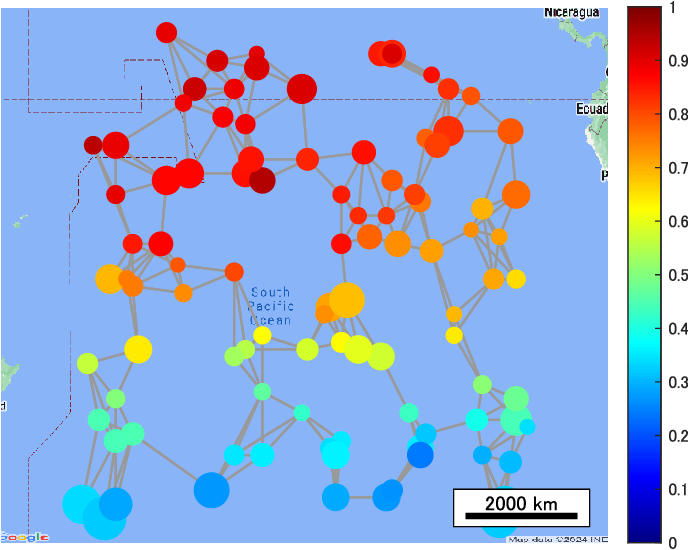}
             RMSE: $ 0.0495$ \\  $\mathrm{L}_2$
             \vspace{2mm}
          \end{center}
        \end{minipage}

         \begin{minipage}{0.20\hsize}
          \begin{center}
            \includegraphics[width=\hsize]{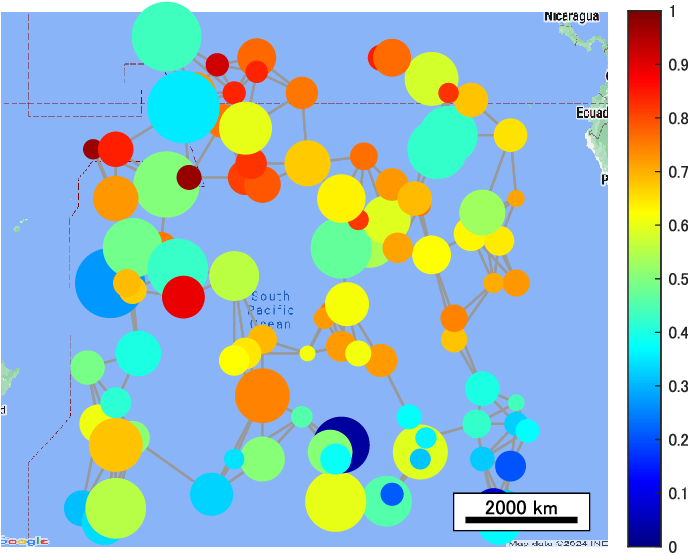}
            \\ RMSE: $ 0.17704$ \\  \textcolor{blue}{S-X} \cite{7117446}
            \vspace{2mm}
          \end{center}
        \end{minipage}
        
    
         \begin{minipage}{0.20\hsize}
          \begin{center}
            \includegraphics[width=\hsize]{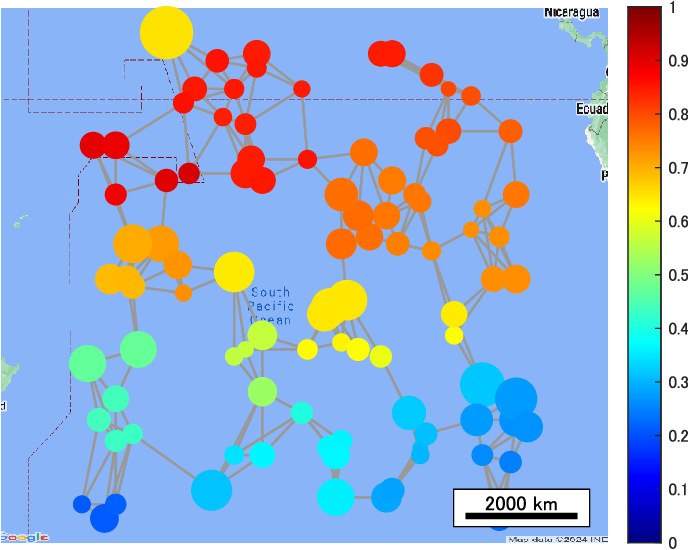}
        \\ RMSE: $ 0.0820$ \\  \textcolor{red}{D-X}
         \vspace{2mm}
          \end{center}
        \end{minipage}

        \begin{minipage}{0.20\hsize}
          \begin{center}
            \includegraphics[width=\hsize]{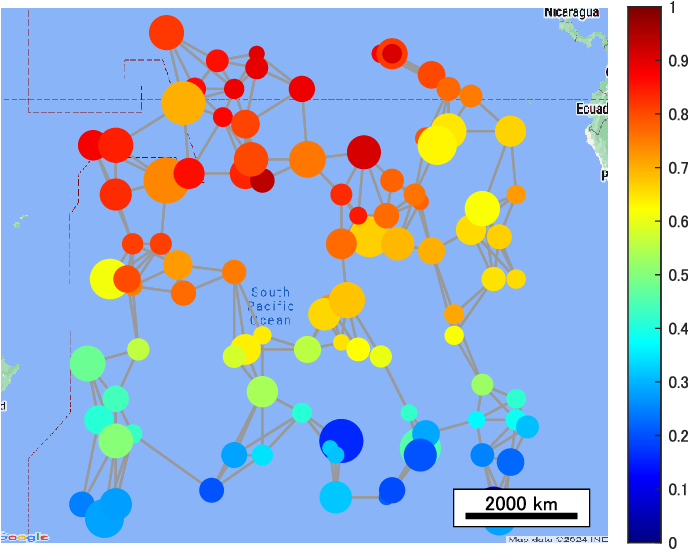}
        \\ RMSE: $ 0.0751$ \\  \textcolor{blue}{TGSR} \cite{7979523}
        \vspace{2mm}
          \end{center}
        \end{minipage}

        \\
        \hspace{-0.21\hsize}
        \begin{minipage}{0.20\hsize}
          \begin{center}
            \includegraphics[width=\hsize]{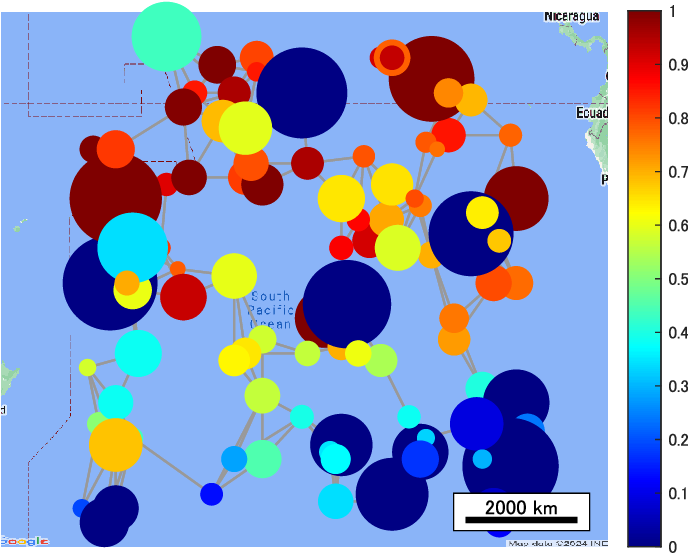}
             \\ RMSE: $ 0.2978$ \\ Obsevation
             \vspace{2mm}
          \end{center}
        \end{minipage}

        \begin{minipage}{0.20\hsize}
          \begin{center}
            \includegraphics[width=\hsize]{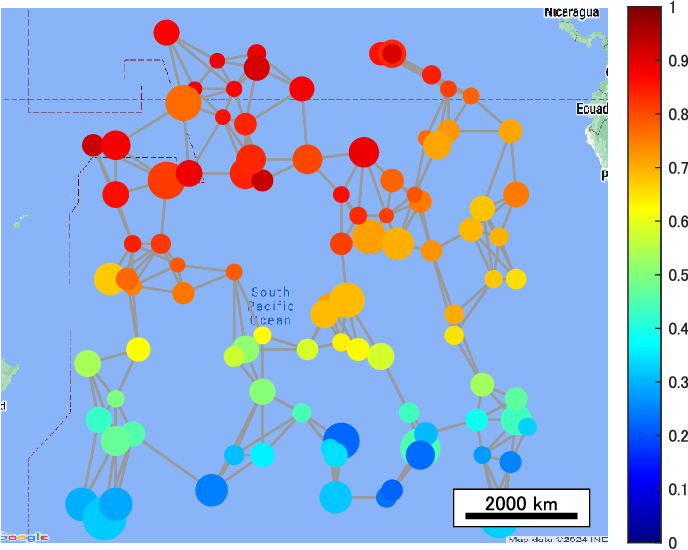}
          \\ RMSE: $ 0.0555$ \\  \textcolor{blue}{GTRSS} \cite{9730033}
          \vspace{2mm}
          \end{center}
        \end{minipage}

        \begin{minipage}{0.20\hsize}
          \begin{center}
            \includegraphics[width=\hsize]{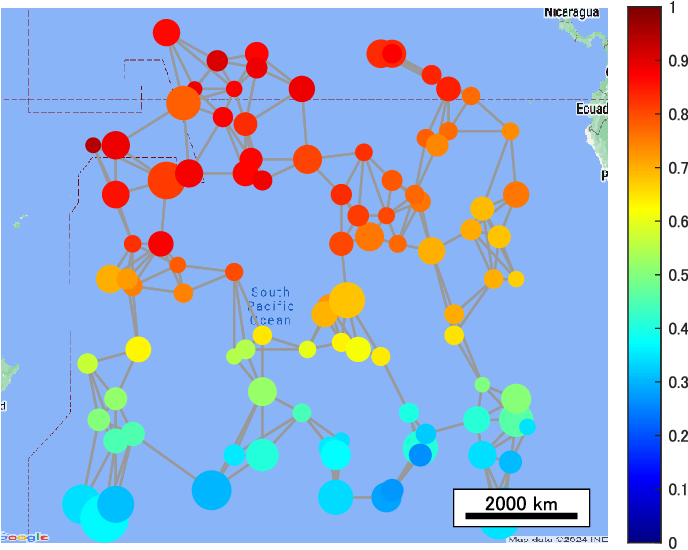}
          \\  RMSE: $ 0.0565$ \\  \textcolor{blue}{S-X-$\mathrm{L}_2$} 
          \vspace{2mm}
          \end{center}
        \end{minipage}
  
        \begin{minipage}{0.20\hsize}
          \begin{center}
            \includegraphics[width=\hsize]{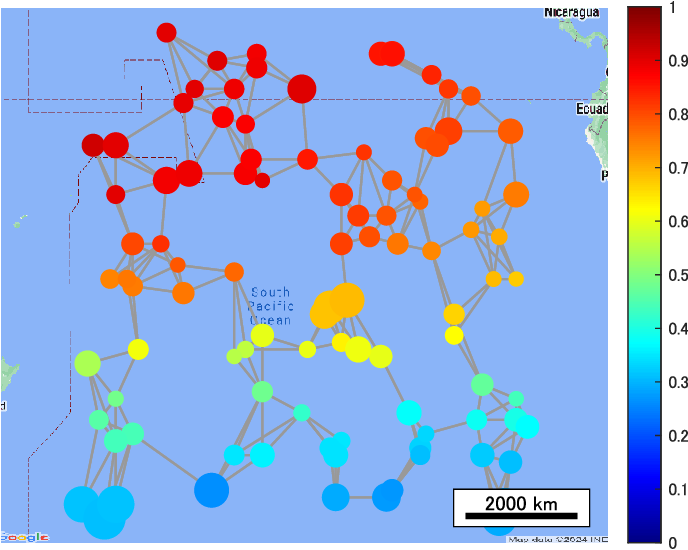}
          \\ {\bf RMSE: ${\bf 0.0428}$ \\  \textcolor{red}{D-X-$\mathrm{L}_2$} Ours}
          \vspace{2mm}
          \end{center}
        \end{minipage}

      \end{tabular}
      \caption{Graph signal recovery results of the sea surface temperature data (Pacific Ocean) ($\sigma,P_s,P_p = 0.1$). The area of the nodes is proportional to the magnitude of the error, where larger areas indicate larger errors.
      Note that although the color range for the observations is clipped to [0,1] to unify the colorbars of all the figures, the magnitude of error represented by the size of the nodes is not clipped, therefore, some nodes may be colored the same but differently sized.} 
      \label{fig:denoise_real}
    \end{center}
  \end{figure*}

\subsection{Results and Discussion}
\label{sec:results}
\subsubsection{Dynamic vs Static Graph Laplacian}

Looking at the results in Table \ref{tb:result} and Figs. \ref{fig:ablation} and \ref{fig:ablation_sst}, the methods that integrate time-varying graph Laplacian-based regularization (the methods in red, namely, Methods D-X, D-DX-$\mathrm{L}_1$, D-DX-$\mathrm{L}_2$, D-X-$\mathrm{L}_1$, D-X-$\mathrm{L}_2$) generally outperform their static counterparts, which only consider a static graph (methods in blue, namely, Methods S-X, S-DX-$\mathrm{L}_1$, S-DX-$\mathrm{L}_2$, S-X-$\mathrm{L}_1$, S-X-$\mathrm{L}_2$). 
In particular, when comparing Methods S-X and D-X, which represent a direct comparison between static and dynamic graph-based regularization, the dynamic regularization consistently shows better performance.

Moreover, comparing Methods D-X-$\mathrm{L}_2$ and S-X-$\mathrm{L}_2$ to Methods $\mathrm{L}_2$ (15th and 10th row to the 2nd row in Table \ref{tb:result}), the static graph Laplacian-based regularization even seems to be hindering the overall recovery performance. This finding highlights that using a constant $\L$ across all time slots should be avoided when the graph is inherently dynamic.

Regarding the methods that employed the regularization $R_v{(\DD(\X))}$ (Methods D-DX-$\mathrm{L}_1$/$\mathrm{L}_2$: 10th and 11th row in Table \ref{tb:result}), the performance was not as strong as initially expected. 
While we hypothesized that the dynamic versions would outperform their static counterparts, the results indicate that the dynamic and static versions perform similarly in most cases. 
We believe this may be due to the fact that the term $\x_{k+1} - \x_k$ is computed using $\L_k$ in the dynamic version, while $\x_{k+1}$ is actually observed on $\L_{k+1}$. This mismatch could result in suboptimal performance. However, we suspect that this effect is dataset-dependent. 

By inspecting the results closely, we observed that the dynamic version does indeed outperform the static version in the sea surface temperature dataset. 
The difference in performance between the synthetic and sea surface temperature datasets seems to stem from the level of smoothness in the vertex domain. 
The sea surface temperature dataset exhibits significantly higher smoothness in the vertex domain compared to the synthetic dataset, meaning the difference between $\L_k$ and $\L_{k+1}$ is smaller. 
When the difference between $\L_k$ and $\L_{k+1}$ is smaller, the term $\x_{k+1} - \x_k$ is more likely to exhibit smoothness on $\L_k$, thereby improving the performance of the dynamic version.

Additionally, $R_v{(\DD(\X))}$ (=$(\x_{k+1} - \x_k)^\top\L_k(\x_{k+1} - \x_k))$ was originally designed to address cases where the vertex domain is less smooth than typically assumed, while also capturing temporal characteristics. 
In our experiments (both on synthetic and sea surface temperature datasets), the vertex domain was found to be very smooth, and we explicitly added regularization terms to handle the temporal domain. 
We expect that the performance of $R_v{(\DD(\X))}$ would improve in datasets where the vertex domain is less smooth.

Finally, our method achieved superior results compared to the state-of-the-art static graph-based methods TGSR and GTRSS (6th and 7th row in Table \ref{tb:result}), which further supports the advantage of incorporating dynamic graphs.

\subsubsection{Vertex vs Temporal Domain Priors}
\label{sec:vs}

In this section, we discuss the impact of vertex and temporal domain priors on recovery performance. 
Comparing Methods $\mathrm{L}_1$ and $\mathrm{L}_2$ (which leverage only temporal domain priors, 1st and 2nd row in Table \ref{tb:result}) to Methods S-X and D-X (which leverage only vertex domain priors, 4th and 5th row in Table \ref{tb:result}), we observe that leveraging temporal domain priors is much more effective than leveraging vertex domain priors for the synthetic dataset.
This is further supported by the relatively small performance gains between Methods $\mathrm{L}_1$, $\mathrm{L}_2$, and Methods D-X-$\mathrm{L}_1$, D-X-$\mathrm{L}_2$ (which leverage priors from both domains).
We speculate that this is because the low velocity and noise levels make the graph signal extremely smooth in the temporal domain, whereas the vertex domain is relatively less smooth even in the noiseless case (Fig. \ref{fig:data} (a)).
Therefore, the temporal domain prior is sufficient for signal recovery, and the additional advantage of using the vertex domain prior is relatively small.

However, the benefits of leveraging the vertex domain prior become more apparent in the real-world dataset (4th and 5th row in Table \ref{tb:result}).
This is because the sea surface temperature dataset is much smoother in the vertex domain compared to the synthetic dataset.
Although the difference in performance between the two regularizations (1st, 2nd row and 4th, 5th row in Table \ref{tb:result}) depends on the nature of the dataset (specifically, the smoothness of the signals in their respective domains), the proposed methods can balance the effects of the two regularizations through the parameter $\lambda$.
Considering that our dynamic methods generally performed better, especially in high-noise settings (Figs. \ref{fig:ablation} (a) and \ref{fig:ablation_sst} (a)), our methods demonstrate robustness to the varying nature of datasets.
Supporting this, our method D-X-$\mathrm{L}_2$ (15th row in Table \ref{tb:result}) performed the best in most settings.

\subsubsection{Sensor Velocity}

From Fig. \ref{fig:ablation} (d), we can observe that all methods leveraging the temporal domain show a decline in performance as the velocity $v$ of the sensors increases. 
This decline occurs because, as the velocity increases, the signals become less smooth in the temporal domain. Consequently, the performance of Methods $\mathrm{L}_1$ and $\mathrm{L}_2$ deteriorates, while Method D-X remains largely unaffected. 
Similarly, the performance gap between Methods D-X-$\mathrm{L}_2$ and $\mathrm{L}_2$ (Fig. \ref{fig:ablation} (d)) widens as the velocity increases, further demonstrating the advantage of incorporating dynamic graph-based regularization.

Methods that rely on static graphs also experience a decrease in performance as the sensor velocity increases. 
This is because the higher the sensor velocity, the greater the mismatch between the single static graph these methods use and the actual, time-varying graphs on which the signals are observed. 
Thus, static methods struggle to capture the true dynamics of the signals in high-velocity scenarios.

In practical settings, differences in sensor velocity are often equivalent to variations in sampling frequency. 
A sensor sampling at a constant frequency while moving at different velocities is essentially equivalent to a sensor moving at a constant velocity but sampling at different frequencies, assuming the underlying signal distribution remains time-invariant. 
Our method demonstrates the ability to tolerate higher velocities compared to the temporal domain-only methods $\mathrm{L}_1$ and $\mathrm{L}_2$, meaning it can also tolerate lower sampling frequencies. 
This indicates that leveraging both vertex and temporal domains can lead to a more cost-efficient recovery method by allowing for lower sampling frequencies without a significant drop in performance.

\subsubsection{Noise Levels}

We conducted extensive experiments across various noise levels, including $\sigma$, $P_s$, and $P_p$ (see Fig. \ref{fig:ablation} and Fig. \ref{fig:ablation_sst}). 
As expected, the recovery performance degrades as the noise level increases. 
However, the proposed method consistently outperforms the state-of-the-art time-varying graph signal recovery methods across all noise settings. 
It is important to note that the comparison methods were modified to support recovery from sparse outliers to ensure a fair comparison. 
We anticipate an even larger performance gap when compared to the original, unmodified formulations of these methods.

\subsubsection{Vertex and Edge Density}

As shown in Fig. \ref{fig:graph_size}, methods using graph Laplacian-based regularizations generally perform better as the number of vertices (sensors) increases. 
This is because, in our synthetic setting, a larger number of vertices leads to a spatially denser sensor network. 
We generate the graph Laplacians using the $k$-nearest neighbors algorithm ($k=4$) and Gaussian kernels based on spatial coordinates, under the assumption that vertices in close spatial proximity have similar signal values. 
While graph Laplacians generated in this way are known to provide smooth and accurate representations of graphs in many situations, the quality of the representation can vary depending on the data. 
In our case, a denser graph implies that the four vertices each vertex is connected to are spatially closer and, therefore, more likely to have similar signal values. 
As a result, the setting with more vertices aligns more closely with the assumptions used to generate the data, improving performance.

We also investigated the impact of edge density on the sea surface temperature dataset (Fig. \ref{fig:ablation_sst} (d)). 
Although performance varies slightly depending on the value of $k$ used in the $k$-nearest neighbor algorithm to generate the graph Laplacians, the overall trend and performance of the proposed method remain consistent. 
Thus, we can confirm that the proposed method is robust to the choice of $k$, and the standard coordinate-to-graph generation scheme is sufficient to support graph-based signal recovery.

\begin{table}[t]
  \caption{The Average Running Time for Method D-X Implemented With ADMM And PDS}
  \label{tb:time}
  \centering 
  \begin{tabular}{ccc}
    \toprule
    \multirow{2}{*}{Dataset size} & \multicolumn{2}{c}{Time (sec.) / per iteration ($\times 10^{-4}$ sec.)} \\
    \cmidrule{2-3} 
    & ADMM & PDS \\
    \midrule
    $64\times100$ & 0.3137 / 4.140 & 0.1118 / 4.395 \\
    $64\times200$ & 0.6776 / 8.846 & 0.2688 / 8.889 \\
    $128\times100$ & 1.0780 / 18.40 & 0.6297 / 18.21 \\
    $128\times200$ & 2.4727 / 42.38 & 1.4246 / 41.47 \\
    $256\times100$ & 4.3046 / 93.89 & 3.3033 / 81.89 \\
    $256\times200$ & 8.6370 / 181.1 & 6.2086 / 158.5 \\
    $512\times100$ & 45.194 / 604.2 & 30.144 / 584.1\\
    $1024\times100$ & 147.12 / 2057 & 127.31 / 2024 \\
    \bottomrule
  \end{tabular}
\end{table}

\subsubsection{Smooth and Piece-wise Flat Signals}

For methods that leverage the temporal prior, we initially expected those using Frobenius-norm regularization to perform better on smooth datasets, and methods using the $\|\cdot\|_1$ regularization term to excel on piece-wise flat datasets, given that $\|\cdot\|_1$ regularization typically performs better on data with staircase-like structures. 
However, contrary to our expectations, methods using $\|\cdot\|_1$ and $\|\cdot\|_F^2$ regularizations performed comparably on the piece-wise flat dataset (1st and 2nd row in Table \ref{tb:result}). 

We speculate that this outcome is due to high noise levels and sensor velocity, which likely disrupted the sparsity of the temporal difference signal. 
As a result, the smoothing ability of Frobenius-norm regularization may have outweighed the advantages of $\|\cdot\|_1$ regularization in this scenario.

\subsubsection{Visual Analysis}

Referring to the visual results of the experiments (Figs. \ref{fig:denoise_128_} and \ref{fig:denoise_real}), we can observe that our methods are able to recover the graph signals fairly well. 
In these figures, the area of each node is proportional to the magnitude of the error, with larger areas indicating larger errors. 
It is important to note that, while the color range for the observations is clipped to [0,1] to standardize the color bars across all figures, the magnitude of the error represented by the node sizes is not clipped. 
As a result, some nodes may have the same color but different sizes, indicating different error magnitudes.

In Fig. \ref{fig:denoise_128_}, we see that a small region in the top middle of the graph, where signal values are high, is being overly smoothed. 
Beyond the challenge posed by the high noise level, the underlying distribution characteristics (Fig. \ref{fig:data}) also contribute to the difficulty of the problem. 
The part of the distribution with high values manifests as spike signals in both the vertex and temporal domains, making accurate recovery harder.

For the same reason, lower sensor velocities help preserve signal smoothness in the temporal domain, thereby improving the performance of methods that leverage temporal domain priors.

\subsubsection{Algorithm Efficency}

We present the computational time of a PDS-based algorithm in comparison to a commonly used Alternating Direction Method of Multipliers (ADMM)-based algorithm \cite{admm}. 
Method D-X was reimplemented using ADMM, and the running times are compared in Table \ref{tb:time} (dataset: synthetic, $\sigma$, $P_s$, $P_p = 0.5$). 
Unlike ADMM, the PDS algorithm avoids matrix inversions, leading to faster computation times. 
This difference becomes more pronounced in the time per iteration as the data dimension increases, since matrix inversions for larger data sizes can be computationally complex.

Additionally, we provide a figure showing the convergence of the PDS-based algorithm in Fig. \ref{fig:convergence}. 
It is important to note that the comparison methods were reformulated and reimplemented to handle the same constraints as our method, ensuring a fair comparison.

\section{Conclusion}

In this paper, we proposed a novel time-varying graph signal recovery problem based on an explicitly dynamic graph model, where the underlying graph is assumed to vary over time, similar to the signals. 
Our formulation incorporates time-varying graph Laplacian-based regularization and temporal difference-based regularization, effectively leveraging priors from both the vertex and temporal domains, while also accounting for the dynamic nature of the graph. 
By jointly estimating sparsely modeled outliers with the graph signals, our method achieves robustness against various types of noise.

Through experiments on both synthetic and real-world data, we compared our method with existing approaches that apply regularization in the vertex and temporal domains. 
We discussed how the priors from both domains contribute to recovery performance under different conditions. 
The proposed problem setting is practical, especially in applications such as remote sensing using mobile sensors, where dynamic graph Laplacians can be easily generated to represent spatial correlations between sensors. 
Dynamic sensor networks offer a practical and feasible approach to sensing, and we view our research as a foundational step toward addressing this emerging problem.

As for possible steps research in this field could take in the future, we suggest the following:
\begin{itemize}
  \item Addressing situations where graphs are unavailable or cannot be easily generated.
  \item Handling cases where the number of vertices varies over time.
  \item Leveraging the priors of adjacent graphs to further enhance signal recovery performance.
\end{itemize}

\ifCLASSOPTIONcaptionsoff
  \newpage
\fi

\begin{IEEEbiography}[{\includegraphics[width=1in,height=1.25in,clip,keepaspectratio]{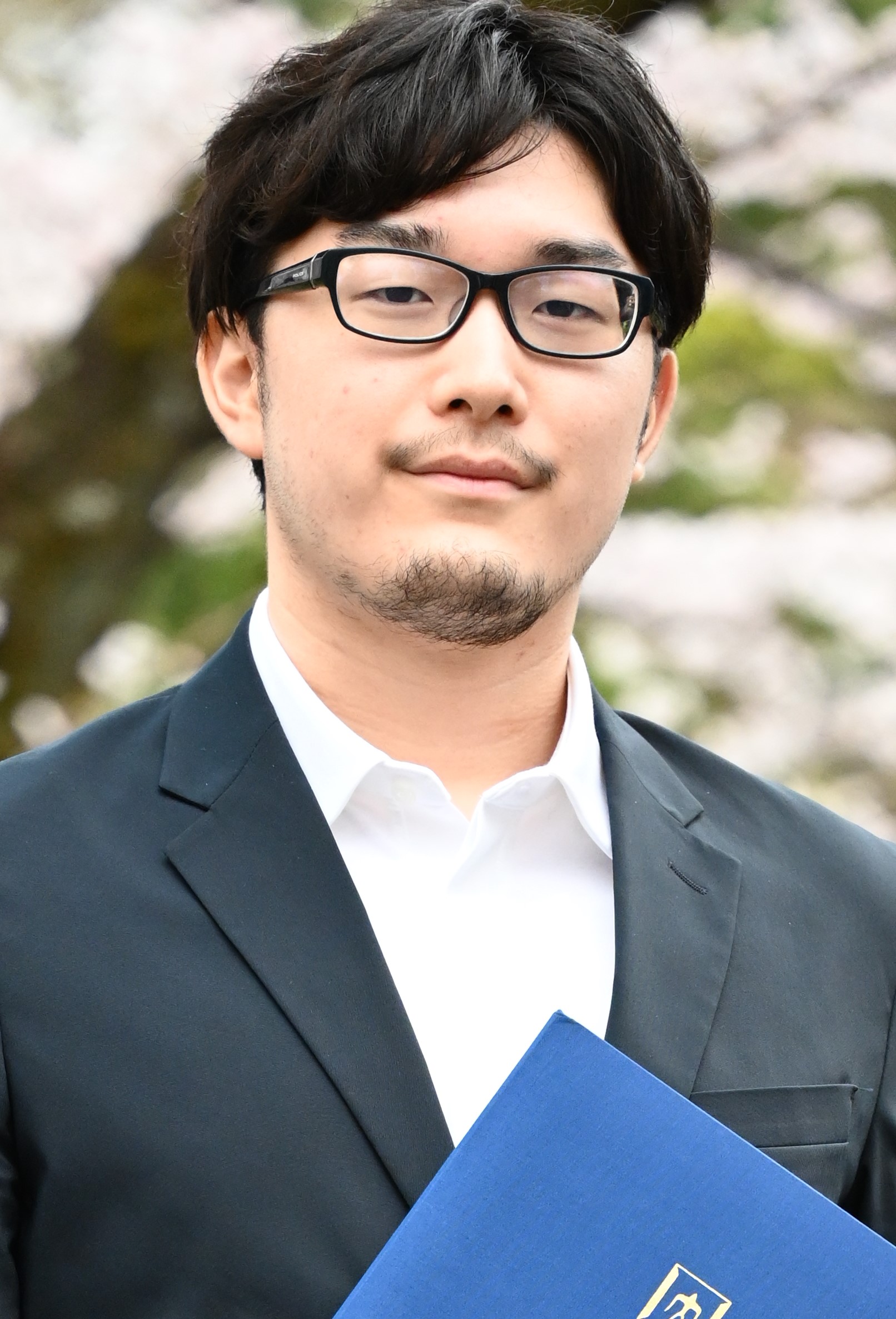}}]{Eisuke Yamagata}
  Eisuke Yamagata (S’21) received a B.E. and M.E degrees in Computer Science in 2020 and 2022 from the Tokyo Institute of Technology, respectively.
  He is currently pursuing an Ph.D. degree at the Department of Computer Science in the Tokyo Institute of Technology. His current research interests are in graph signal processing and time series analysis. Since October 2023, he has been a Research Fellow (DC2) of Japan Society for the Promotion of Science (JSPS).
  \end{IEEEbiography}

  \begin{IEEEbiography}[{\includegraphics[width=1in,height=1.25in,clip,keepaspectratio]{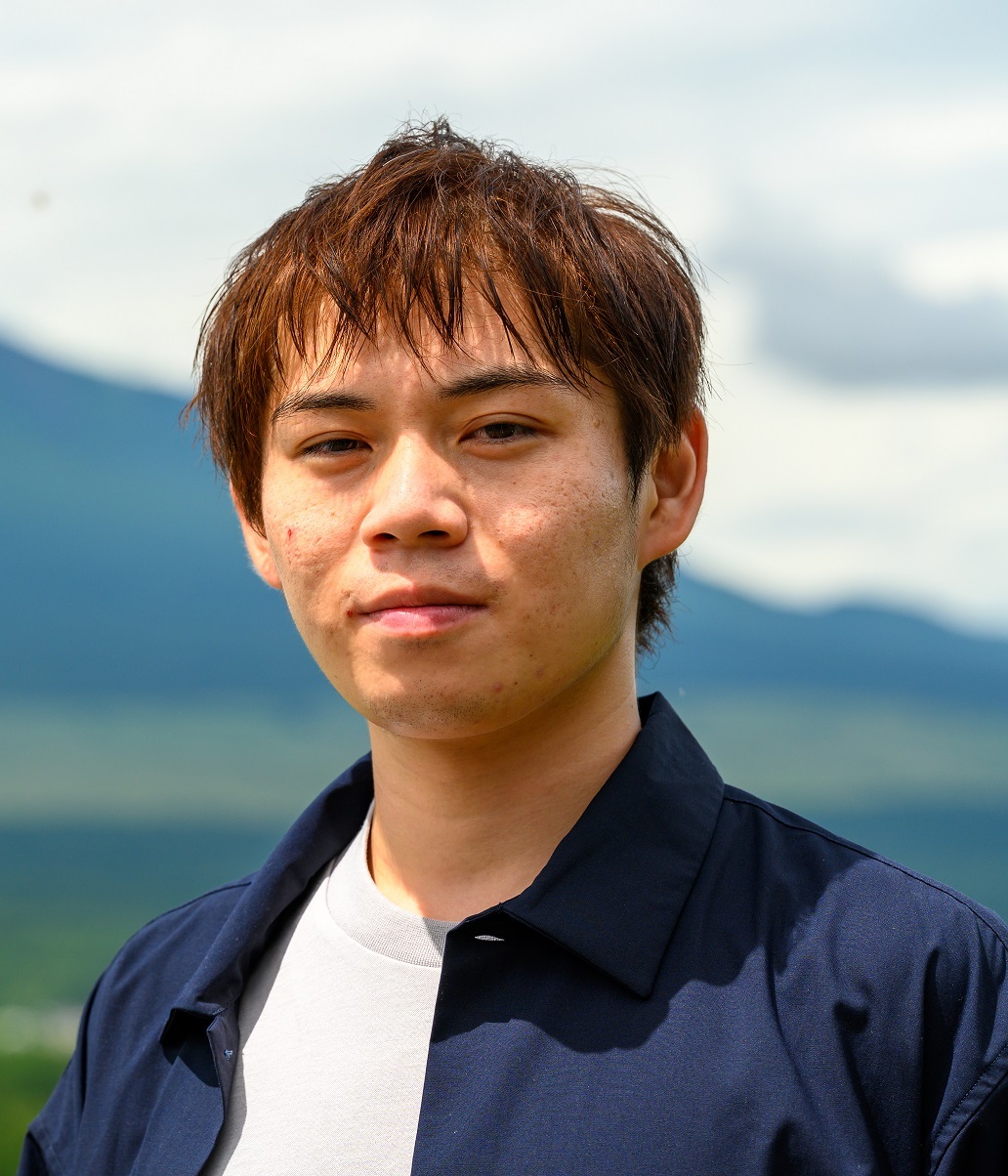}}]{Kazuki Naganuma}
  Kazuki Naganuma (S’21) received the B.E. degree in information and computer sciences from Kanagawa Institute of Technology, Atsugi, Japan, in 2020, and the M.E. degree and Ph.D. degree in information and computer sciences from Tokyo Institute of Technology, Yokohama, Japan, in 2022, where he is currently a researcher at Institute of Science Tokyo. Since October 2023, he has been a Research Fellow (DC2) of Japan Society for the Promotion of Science (JSPS) and a Researcher of ACT-X of Japan Science and Technology Agency (JST), Tokyo, Japan. His research interests include signal and image processing and optimization theory. Dr. Naganuma received the Student Conference Paper Award from the IEEE SPS Japan Chapter in 2023.
  
  \end{IEEEbiography}

  \begin{IEEEbiography}[{\includegraphics[width=1in,height=1.25in,clip,keepaspectratio]{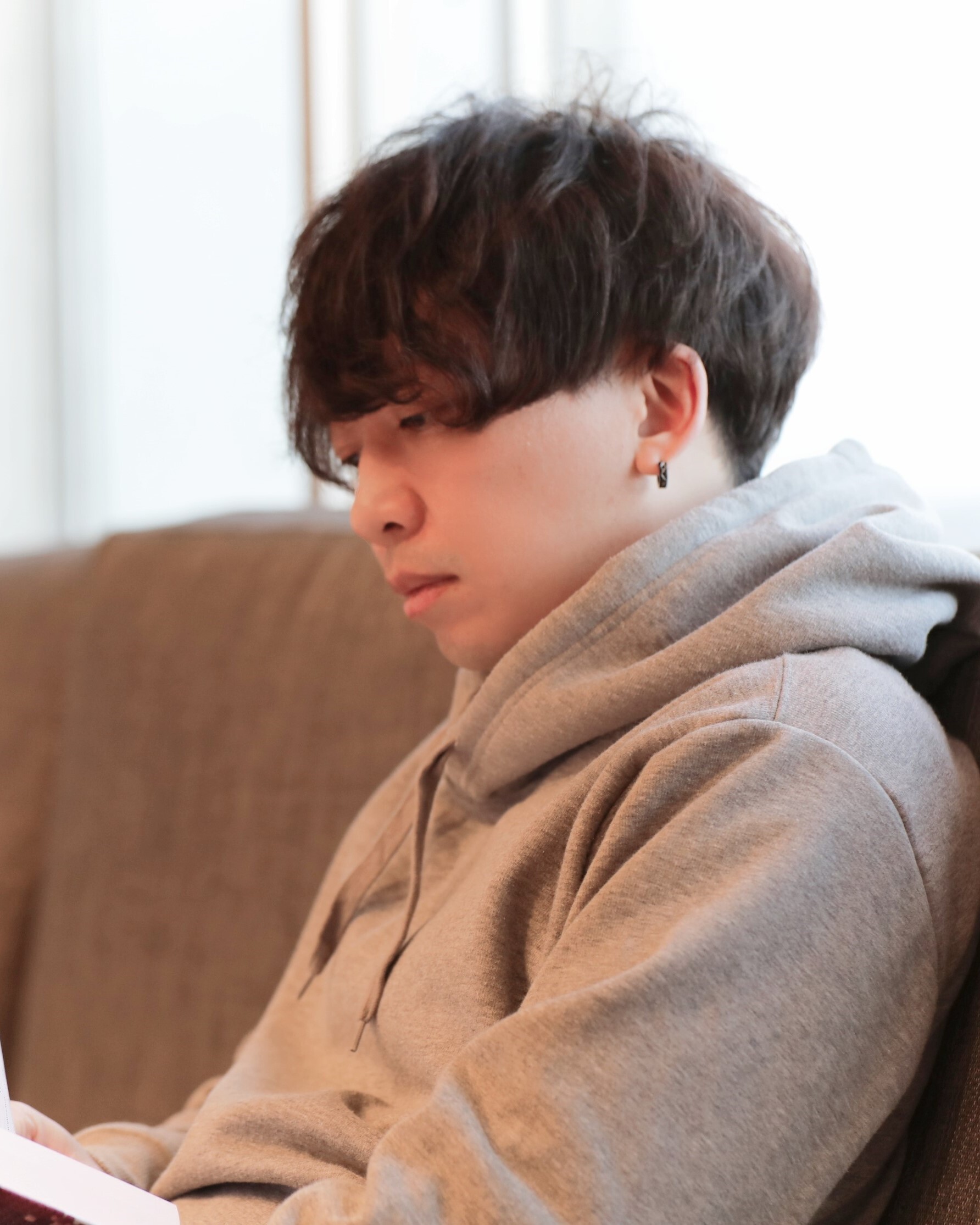}}]{Shunsuke Ono}

  Shunsuke Ono (S’11–M’15–SM'23) received a B.E. degree in Computer Science in 2010 and M.E. and Ph.D. degrees in Communications and Computer Engineering in 2012 and 2014 from the Tokyo Institute of Technology, respectively.

From April 2012 to September 2014, he was a Research Fellow (DC1) of the Japan Society for the Promotion of Science (JSPS). He is currently an Associate Professor in the Department of Computer Science, School of Computing, Tokyo Institute of Technology. From October 2016 to March 2020 and from October 2021 to present, he was/is a Researcher of Precursory Research for Embryonic Science and Technology (PRESTO), Japan Science and Technology Agency (JST), Tokyo, Japan. His research interests include signal processing, image analysis, remote sensing, mathematical optimization, and data science.

Dr. Ono received the Young Researchers’ Award and the Excellent Paper Award from the IEICE in 2013 and 2014, respectively, the Outstanding Student Journal Paper Award and the Young Author Best Paper Award from the IEEE SPS Japan Chapter in 2014 and 2020, respectively, the Funai Research Award from the Funai Foundation in 2017, the Ando Incentive Prize from the Foundation of Ando Laboratory in 2021, the Young Scientists’ Award from MEXT in 2022, and the Outstanding Editorial Board Member Award from IEEE SPS in 2023. He has been an Associate Editor of IEEE TRANSACTIONS ON SIGNAL AND INFORMATION PROCESSING OVER NETWORKS since 2019.
  \end{IEEEbiography}



\end{document}